# Long-Propagating Ghost Phonon Polaritons Enabled by Selective Mode Excitation.

*Manuka P. Suriyage* [1#]*, Qingyi Zhou*[2#]*, Hao Qin*[1]*, Xueqian Sun*[1]*, Zhuoyuan Lu*[1]*, Stefan A. Maier*[4]
*, Zongfu Yu*[2*] *and Yuerui Lu*[1,3*]

[1]School of Engineering, College of Engineering, Computing & Cybernetics, the Australian National University, Canberra, ACT, 2601, Australia

[2]Department of Electrical and Computer Engineering, University of Wisconsin-Madison, Madison, WI 53706, USA

[3]ARC Centre of Excellence in Quantum Computation and Communication Technology ANU node, Canberra, ACT 2601, Australia

[4]School of Physics and Astronomy, Monash University Clayton Campus, Melbourne, Victoria 3800, Australia

# Those two authors contribute equally to this work.

ABSTRACT

The precise control of phonon polaritons(PhPs) is essential for advancements in nanophotonic applications like on-chip optical communication and quantum information processing. Ghost hyperbolic phonon polaritons (g-HPs), which have been recently discovered, feature in-plane hyperbolic dispersion and oblique wavefronts, enabling long-range propagation. Despite their potential, controlling the directionality and selective excitation of g-HPs remains challenging. Our research demonstrates that modifying the shape of the launching micro/nano antenna can achieve this control. Using an asymmetric triangular gold antenna on a calcite crystal surface, we achieve highly directional g-HP excitation by selectively targeting specific polariton modes. Additionally, the mode of g-HPs can be adjusted by changing the excitation wavelength or rotating the antenna. Remarkably, our near-field imaging experiments show g-HP propagation over distances exceeding 35 micrometers, more than twice the length reported in previous studies. This work merges g-HP theory with structural engineering, enhancing the control over g-HPs and paving the way for innovative applications in mid-IR optoelectronics.

**Introduction**

Optical anisotropy is a phenomenon where the dielectric permittivity varies resulting in polarized light propagating with distinct velocities and wavelengths along different crystal axes[1,2]. Natural crystals including calcite have been employed for extensive applications. Their dielectric function is also constant over a broad spectral range[3] facilitating design of functional components. In such crystals, a huge difference in the refractive index along orthogonal axes can be observed, enhancing the material anisotropy[4]. In some cases, the permittivity tensor can become negative within a spectral range allowing for the excitation of surface polaritons[5]. Natural crystals can be used to produce hyperbolic phonon polaritons if the permittivity tensor elements have opposite signs in principal axes $\varepsilon_t \varepsilon_z < 0$ ( $\varepsilon_t = \varepsilon_x = \varepsilon_y$ denote the in-plane components, $\varepsilon_z$ denotes the vertical component). In such systems, the isofrequency curve of light is an open hyperbola[5]. Numerous materials that carry both in-plane and out-of-plane hyperbolic polaritons have been investigated[6], from van der Waals thin films[2,7] which supports a single optical axis and crystallize in hexagonal[8,9], trigonal[10], and tetragonal[11] systems to biaxial materials with two optical axes in orthorhombic[12,13], monoclinic[14], and triclinic materials. Two main types of hyperbolic polaritons, namely volume-confined hyperbolic polaritons[15] and surface-confined hyperbolic polaritons[16], have been investigated[17]. Most recently, g-HPs[18] were discovered, which exhibits in-plane hyperbolic dispersion on the surface of a polar uniaxial crystal as well as oblique wave fronts in the bulk, similar to recently predicted ghost waves[19]. These g-HPs showcase a complex out-of-plane wavevector within the crystal even in the absence of material loss[18]. Due to the oblique crystal lattice and the ability to control the optical axis angle with respect to the material surface (through mechanically cutting and polishing the sample), calcite offers precise control over its polaritonic response. This ability to finely adjust the optical axis enables a hyperbolic to elliptic topological transition of the polariton waves in calcite substrates. Based on this recent discovery[18], the optic angles $\theta = 23°$ (surface cut along [100] plane), $\theta = 90°$ (surface cut along [001] plane) effectively rotates the IFC (iso-frequency curve) in k space, resulting in the generation of g-HPs (Supporting Information Note 1) only below the critical angle of $\theta = 57°$.

Control over the excitation and propagation of phonon polaritons allow useful opportunities for many nanophotonic applications[20-26]. The ability to achieve selective excitation with high directionality and long-propagating PhPs is of particular interest. Highly directional propagation of PhPs is crucial for applications such as optical communication[27], quantum information processing[28], coupling between quantum emitters[29] and heat management[30,31]. In addition, long propagation distance is highly desirable for achieving high signal-to-noise ratio and low-loss transmission. Both of these issues could be solved if it is possible to selectively excite polariton modes that contain the desired properties. However, the current state-of-the-art techniques face significant limitations in achieving both highly directional and long-propagating PhPs simultaneously[32-34].

There are multiple approaches to excite phonon polaritons, which involve utilizing the physical edge of the polariton-supporting material[35], creating a difference in the dielectric environment below and/or above the 2D polaritonic medium[5], and introducing a highly scattering object, such as a metal nanostructure or nano antennas, in order to quasi-phase-match[36]. The combination of scattering-type scanning near field optical microscope (s-SNOM)[37] and resonant micro/nano antennas is frequently used to directly probe phonon polaritons. Tip-induced excitations can be eliminated by employing a dielectric AFM (Atomic Force Microscope) tip.

In this paper, we present a novel approach to achieve selective mode excitation with highly directional and long-propagating PhPs by utilizing an asymmetric micro/nano antenna on a calcite surface. We demonstrate that the directionality and propagation length of PhPs can be flexibly controlled by varying the antenna's shape, excitation wavelength, and orientation of the antenna. The microstructure of the triangular shaped asymmetric antenna is analyzed deeply to provide a technique to selectively excite the polariton modes. Our near-field imaging experiments reveal that the PhPs excited by the triangular antenna can propagate over a very long distance (>35 micrometers), much longer than the results reported in literatures[12,18]. This breakthrough is achieved by combining g-HP theory with structural engineering, which allows us to only excite certain g-HP modes and overcome the limitations of obtaining desired polariton properties such as high directionality and larger propagation lengths. Overall, our work provides new insights into the underlying physics of PhPs and opens up opportunities for various applications. The technique introduced here is general and can be extended to other anisotropic materials and other types of polariton modes.

This paper is structured as follows: initially, we discuss the innovative design strategies employed to induce selective excitation of polariton modes in highly directional g-HPs, achieved through the utilization of variously shaped micro/nano antennas. Subsequently, we present the investigation of the influence of excitation frequency and antenna shape on the polariton mode excited and the resultant directionality. Following this, we discuss edge-assisted mode selection by considering various shapes and orientations of triangle antennas relative to the crystal axis. After that, we demonstrate the ability of the newly developed micro antennas to create considerably longer g-HPs. Finally, we conclude with a comparison of similar work highlighting the uniqueness of our work and outlook.

**Results and Discussion**

**Shape-dependent selective mode excitation of g-HPs with micro/nano antennas.**

The directionality of a polariton can be crucial for various applications[38,39] where the direction of the electromagnetic flow at nanoscale becomes important. Based on previous studies this has been achieved through grating diffraction which enables the propagation of the polariton only on one side of the grating[39,40], via topological transition in background surrounding media such as photonic crystals[41,42] and hybrid plasmonic structures[14] and most recently through polarized dipoles[43,44] and polaritonic crystals[45]. While the grating technique is widely used in photonics, the resulting unidirectional polaritons generated through gratings differ significantly from the wavefronts launched by a point source, such as a nano antenna[39]. But the most recent discovery of asymmetric propagation involves the utilization of a small disk antenna with varying in-plane polarizations, achieved by manipulating the in plane direction of incident light[43,44]. This is an effective way to break the symmetry of the g-HPs but lacks the ability to selectively excite modes with a broad controllability as we have demonstrated here. Through our experimental demonstration, we utilized different micro antenna shapes to produce highly directional excitations of g-HPs while maintaining the concave wavefronts produced by conventional disk-shaped nano antennas. The motivation comes from the fact that while the calcite surface supports g-HP modes propagating in different directions, it is possible to excite certain modes while suppressing others by merely adjusting the antenna shape. Therefore, by introducing an asymmetric triangular antenna, we demonstrate that only the upper branch g-HP can be strongly excited, leading to a radiation pattern with a desired high directivity. More specifically, three different shapes of antennas were fabricated on the calcite sample (see methods and fabrication section) and the real space imaging of the diffraction of g-HPs was performed using a s-SNOM. Often PhPs are launched more efficiently by the s-SNOM tip than by the sample itself[39]. The experimental observations (Figure 1a-c) match perfectly with the simulation results for the g-HPs

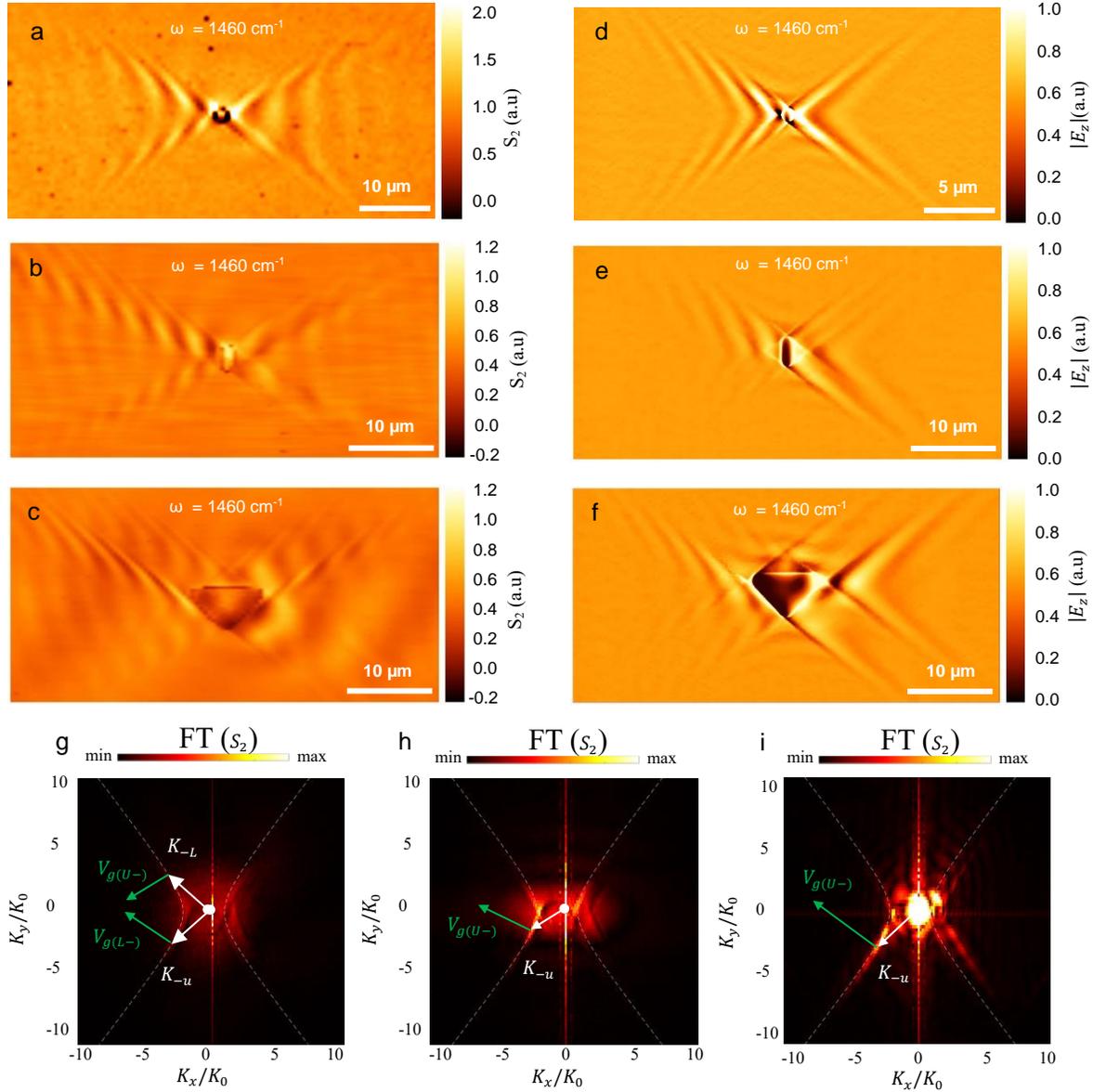

**Figure 1. Demonstration of the selective mode excitation for directional propagation of g-HPs by changing the shape of the micro/nano antenna. a-c,** Experimental near-field images of antenna-launched g-HPs at the illumination frequency $\omega = 1460\ cm^{-1}$. **d-f,** Simulated near-field images of antenna-launched (disk, rectangle, triangle) g-HPs. **g-i,** Absolute value of the Fourier transform of the images shown in a-c, the bright spots represent the modes excited, and the white arrows indicate the wave vectors (momenta) of the antenna-launched g-HPs. Green

arrows indicate the energy flow direction / group velocity $V_{g,(i)}$ of the respective polaritonic branch. White dotted curve lines represent the IFC (left and right).

(Figure 1d-f; generated without considering the s-SNOM tip; Supporting Information Note 2) confirming the absence of tip-launched PhPs in any of our experimental results. We fabricated gold antennas in the shape of disk (diameter D=1.5 μm; Figure S3b), rectangle (H × W = 3.2 × 1.4 μm; Figure S3e), and triangle (H × L = 4.9 × 8.4 μm; Figure S3h) to demonstrate the controllability of the highly directional propagation of the g-HPs on calcite (Supporting Note 3 details the design features of the antennas). The micro/nano antenna concentrates the 60-degree oblique incident illumination (at 1460 cm$^{-1}$) into a localized hotspot or multiple hotspots depending on the design of the antenna (see Figure S2 and S6), which launches the g-HP. The upper and lower polariton branches exhibit varying intensities that demonstrate the selective mode excitation of the polaritons produced by different shaped antennas.

The directionality of the polariton waves was initially confirmed based on the calculated β - directionality constant (equation 1) values and it is further justified by analyzing the FFT results of the respective polaritons.

$$\beta = \left| \frac{amplitude\ of\ the\ upper\ polariton\ branch}{amplitide\ of\ the\ lower\ polariton\ branch} \right| \quad (1)$$

The triangle shape showcases the highest directionality with a β value of 4.35, meaning that the excited U- branch of the g-HP is more than 4 times stronger than the L- branch. In contrast, the disk exhibits the lowest directionality with a β value of 1.05, meaning that the intensity of U- and L- branches of the g-HPs is almost identical. Based on the relationship in (Eq S11; Supporting Information Note 3), the directionality of the g-HP increases from disk, rectangle to the triangle-shaped antenna. The numerical simulation results (Figure 1d-f) show similar results to the experimental measurements, and the directionality varies similarly based on the calculated β values and FFT results.

It is evident that the shape of the micro/nano antenna can selectively excite specific polariton modes. As we can observe in figure 1i the triangular shaped antenna is exciting specific set of polaritonic modes efficiently compared to the other modes. It is not only a single mode that is being excited but a collection of few modes that superimpose creating a polaritonic branch with a specific group velocity (Supporting Information Note 3 and Figure S3). The super positioned polariton's group velocity $V_{g(i)}$ (where i = U+, U-, L+, L-; L: lower branch, U: upper branch) represents the energy flow direction. Figure 1g represents the FFT results of the experimental SNOM measurement in Figure 1a for a disk-shaped antenna. All four polariton branches are visible in the experimental results with similar intensities for the left side branch and varying intensities for the right-side branch. These slightly varying intensities of the polaritonic branches for a disk antenna are present due to the oblique incident, and this is further explained in Supporting Information Note 3. The FT shown in figure 1g showcases that the disk antenna is capable of exciting multiple modes in all 4 branches ( U+, U-, L+, L-) with different coupling efficiencies. Two modes (highest k modes excited by the disk antenna) are visualized using the k vectors ($K_{L-}$, $K_{U-}$) in the FT image with the corresponding group velocities. The brightness of those bright spots in k space represents the intensities of the polaritonic branch with that specific polariton mode. Figure 1h is the FFT result for the polariton generated by the rectangular antenna (Figure 1b). Here the change of shape has modulated the intensity of the U- branch and it is extremely high according to the brightness of the spot that represents, $K_{U-}$ vector in panel h. But still, it is not as efficient as a triangle shape antenna to excite modes selectively. The FFT result (figure 1i) of the polariton generated by the triangular antenna (figure 1c) clearly showcases two bright regions with one having discrete extremely bright spots (U- branch). Therefore, it is clear that the triangular shaped antenna is only exciting a collection of modes that generated two branches with the wavevectors, $K_{U-}$ and, $K_{U+}$ and it is clear that the $V_{g(U-)}$ is higher in intensity compared to the $V_{g(U+)}$ polaritonic branch.

**Edge assisted selective mode excitation.**

Controlling the properties of polaritons is essential in various applications, such as sensing[46], communication, and optoelectronics[23]. This is normally achieved by adjusting the size and refractive index of the polariton system[17,47], modifying the coupling strength[21,48], and using patterned structures[17,18,37]. The excitation frequency of the illumination source determines important parameters of the ghost polariton, particularly the directionality angle α of the hyperbolic polariton (Figure 2c). Our simulation and experimental results show that α increases w.r.t to the illumination frequency, but the rate of increase can be controlled by the shape of the antenna. Specifically, experimental results (Figure 2a-c) demonstrate a linear increase in the directionality angle from 4 to 45 degrees as the illumination frequency is increased from 1420 to 1470 cm$^{-1}$ (Figure 2d). Similarly, simulation results (Figure 2e-g) show that the directionality angle varies from 2 to 45 degrees for the same range of excitation frequency (Figure 2h). This is achieved through selectively exciting polaritonic modes by the antenna. Specifically, the triangular-shaped antenna exhibits reduced opening angles when the excitation wavenumber is below 1470 cm$^{-1}$. This is due to the influence of the polariton generated by the triangular antenna, which is precisely managed through the selective excitation of modes using the curved vertices and edges of the antenna  (Supporting Information Note 3). The triangular antenna's

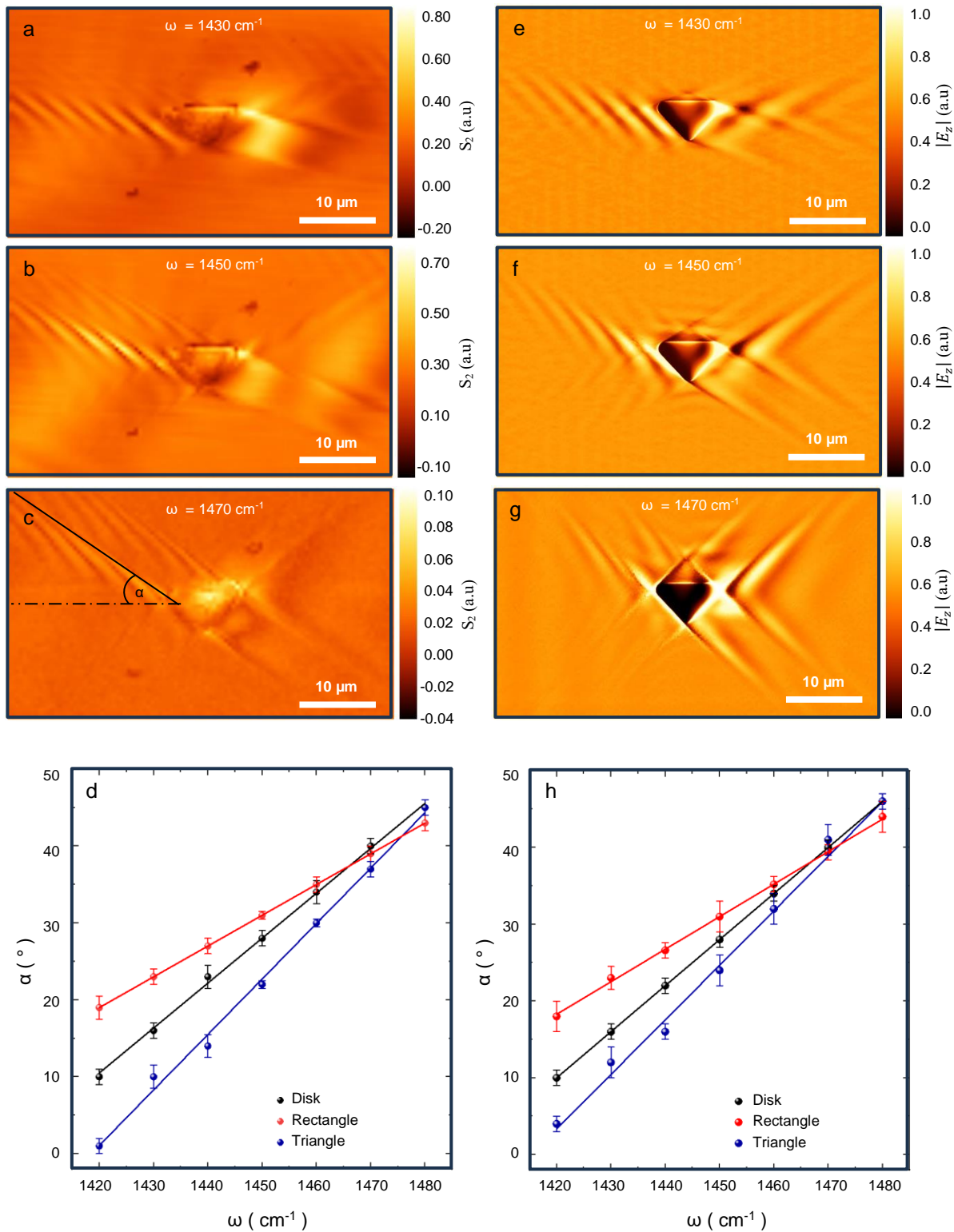

**Figure 2. Edge assisted mode selection and modulation of the directionality angle by the**

**excitation frequency. a-c,** Experimental near-field images of antenna-launched g-HPs at the illumination frequencies of $\omega = 1430\ cm^{-1}$, $\omega = 1450\ cm^{-1}$, $\omega = 1470\ cm^{-1}$. $\alpha$ in panel c is the directionality angle of the excited polariton. **d,** Measured directionality angles of the experimental results for all three different shapes of micro/nano antennas. **e-g,** Simulated near-field images of antenna-launched g-HPs at the illumination frequencies of $\omega = 1430\ cm^{-1}$, $\omega = 1450\ cm^{-1}$, $\omega = 1470\ cm^{-1}$. **f,** Measured directionality angles of the simulation results for all three different shapes of micro/nano antennas. The blue, black and red lines in **d** and **h** are the linear fitting curves for the directionality angle measurements of polaritons excited by triangle, disk and rectangle shaped antennas respectively. The error bars in in **d** and **h** represents the fitting uncertainty.

curved edge triggers a specific set of polaritonic modes towards the U- branch and, simultaneously, the angled physical edge of the triangular antenna facilitates high coupling efficiency to the polariton modes which has wave vectors perpendicular to the edge. Consequently, the polaritons observed through experiments represent an interference pattern resulting from the combination of these effects, resulting in a slightly narrower directionality angle compared to the polaritons excited by a disk micro/nano antenna. Edge assisted mode selection does not have a significant impact on larger wavenumbers because the angle of the physical edge of the triangle antenna is insufficient to couple and scatter modes with wave vectors perpendicular to the physical edge. Therefore, beyond a certain frequency, the directionality angle changes for a triangular-shaped antenna, as the polariton mode is solely determined by the selective excitation of the triangular antenna through the curvatures of the vertices. Therefore, if there is a need to modulate the directionality angle without changing the illumination frequency, using a distinct micro/nano antenna shape can serve the purpose.

To verify these findings triangular antennas were fabricated on calcite with different shapes, sizes and four different orientations relative to the [100] surface. The observed polariton fringes in Figure 2a, b are parallel to the AC physical edge of the triangular antenna. But for the same triangle shape antenna at higher illumination frequencies (Figure 1c and Figure 2c) the polariton branches showcase oblique wavefronts. This occurs due to the edge assisted mode selection which provides a higher controllability in selectively exciting polariton modes (Supporting Information Note 4). By adjusting the internal angle of triangle $\delta$ (as shown in Figure 3a), we can alter the inclination of the AC physical edge of the triangle shape, offering a means to regulate selective mode excitation. The antenna itself excites the polariton and the mode selection will be affected by the vertices (A, B, C) and the AC physical edge of the triangular shaped micro antenna (detailed in Supporting Information Note 4). Even though the IFC supports an infinite number of modes, depending on the curvatures of the three vertices only a few discrete numbers of modes would be efficiently coupled by the antenna to excite polaritons. The highlighted region (orange shaded region of the IFC in Figure 3a) is to showcase that this is limited and can be tailored by changing the curvature and the size of the antenna. If the curved edges (B and C) of the blue color triangle shown in the schematic (Figure 3a) with an internal angle $\delta_1$ can excite the mode, $M_{e1}$ then the physical edge enhances the coupling efficiency of that specific mode. Then if the angle is reduced to $\delta_2$ for the edge to enhance the coupling efficiency the vertices should efficiently excite mode $M_{e2}$. Therefore, when the triangle internal angle $\delta$ reduces the edge assisted coupling mode will have a high k vector. These wave vectors could be coupled efficiently by adjusting the size of the antenna with respect to the curvatures and the edge lengths. But beyond a certain critical angle the edge will not be able to support mode selection as demonstrated through the smallest triangle with an internal angle of $\delta_3$. Now there are no modes supported by calcite medium which has a k vector perpendicular to the AC physical edge of the triangle. But still if needed, we can get parallel fringes to the AC edge of the smallest triangle (with δ=$\delta_3$) by changing the excitation frequency. If we decrease the excitation frequency, the IFC will open, allowing the triangle with an inclined surface at a $\delta_3$ angle to support certain modes, provided the edges can excite the correct polariton mode.

The near field results and the corresponding FFT images in Figure 3b-e illustrate how this physical edge can selectively excite modes. Notably, at $\omega = 1460\ cm^{-1}$, the triangle is exciting multiple modes, as evidenced by the presence of numerous bright spots in the FT image in Figure 3c. The two most prominent excited modes are labelled as $M_{1T}$ and $M_{2T}$ while no modes are activated by the vertices, corresponding to mode $M_i$ which is positioned on the IFC where it intersects with line $l_t$ (perpendicular axis to the AC edge of the triangle). Thus, at $\omega = 1460\ cm^{-1}$ the triangular shape facilitates the coupling of multiple modes with high directionality in the U-branch. Conversely, when the same antenna is coupled with a different illumination wave at $\omega = 1430\ cm^{-1}$ we observe parallel polaritonic fringes, as shown in Figure 3d. The FFT result displayed in Figure 3e reveals a solitary bright spot, indicating the excitation of a single polariton mode. As anticipated, this mode $m_i$ corresponds to the intersection point of the respective IFC at $\omega = 1430\ cm^{-1}$ and line $l_t$, underscoring our successful realization of selective excitation of polariton modes through the utilization of a physical edge.

**Figure 3. Edge assisted mode selection.** (**a**) Schematic illustration of the edge assisted mode selection at the boundary due to polariton excitation and scattering. Black solid line represents the IFC for calcite at 1480 $cm^{-1}$. Blue, red, and green triangles represent three separate triangles with three distinct angles $\delta = \delta_1, \delta_2, \delta_3$. $l_1, l_2, l_3$ are the three lines normal to the AC edge of each triangle. $M_{e1}, M_{e2}$ are the two modes at the intersection of IFC and $l_1, l_2$ lines. $K_{e1}$ and $K_{e2}$ represent the corresponding wave vectors for modes $M_{e1}$ and $M_{e2}$. $k_i$ represent the wave vector for the incident wave. $V_{gi}, V_{gr}$ are the group velocities / energy flux of the incident and reflected light. (**b**) Near-field amplitude images of a triangular antenna-launched g-HPs at the illumination frequency $\omega = 1460$ cm$^{-1}$. Triangle inner angle $\delta_t = 43°$ and $l_t$ is the normal axis to the AC edge of the triangle. (**c**) Absolute value of the Fourier transforms of the images shown in B. $M_{1t}$ and $M_{2t}$ are the antenna excited modes at $\omega = 1460$ cm$^{-1}$. $M_i$ is the intersection point of line $l_t$ and the IFC at $\omega = 1460$ cm$^{-1}$. (**d**) Near-field amplitude images of a triangular antenna-launched g-HPs at the illumination frequency $\omega = 1430$ cm$^{-1}$. (**e**) Absolute value of the Fourier transforms of the images shown in D. $m_i$ is the intersection point of line $l_t$ and the IFC at $\omega = 1430$ cm$^{-1}$.

Then we explored the change of orientation of the micro/nano antenna to modulate the directionality of the polariton waveform generated in both experimental and simulation results. The orientation angle σ of the antenna relative to the crystal axis (Figure 4f; angle w.r.t the x axis of the calcite sample surface) of calcite is a critical parameter governing the diffraction and the directionality of the g-HP. Figure 4a-d illustrates the measured s-SNOM images of the g-HPs with antenna orientation angles σ =0°, 30°, 60° and 90°, respectively. All our experimental near-field images show excellent agreement with the simulation results (Figure 4e-h), supporting the validity of the observed directionality in experimental measurements. Thus, σ can be considered as a new degree of freedom to modulate the directionality of the g-HPs. Here the position of the vertices and the angle of the physical edge of the triangular antenna changes resulting in selective excitation of polaritonic mode as explained before for the non-rotated triangle.

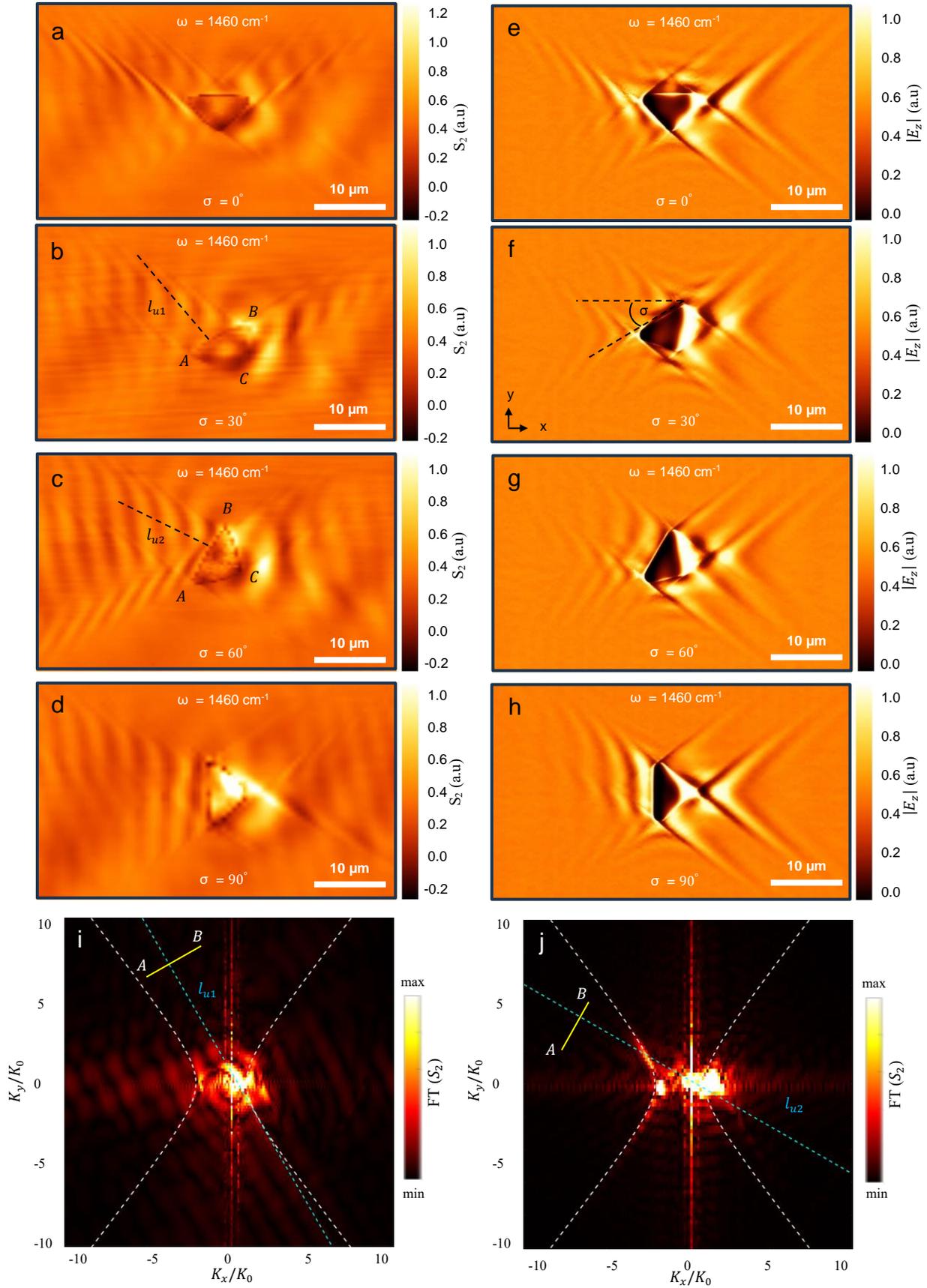

**Figure 4. Edge assisted mode selection based on the rotational angle (σ ) of the micro/nano antenna.** (a-d) Experimental near-field images of antenna-launched g-HPs for the rotational angles σ = 0°, 30°, 60°, 90° . $l_{u1}$ and $l_{u2}$ are the perpendicular axis to the BA physical edge of the triangles with rotational angles σ = 30°, 60°. (e-h) Simulated near-field images of antenna-launched g-HPs, for the rotational angles σ = 0°, 30°, 60°, 90°. (i,j) Absolute value of the Fourier transform of the images B and C. Yellow AB line represents the AB physical edge of the triangle.

Similar to how the AC physical edge facilitated mode selection in the U- branch, we demonstrate the capability to excite modes in the L- polariton branches by rotating the triangle along the crystal axis. As Illustrated in Figure 4, when the triangle is rotated with respect to the x-axis at a 60-degree angle the L- polariton branch is excited, and polariton fringes emerge parallel to the BA (Figure 4c) physical edge. This phenomenon arises from the interplay between polaritons excited by the curved vertex (B) and the physical edge BA, maintaining the parallel fringes to the edge unchanged. The FT result in Figure 4j justifies this, with a bright spot evident at the intersection points of the IFC and the line $l_{u2}$ (the perpendicular axis to the BA edge of a 60-degree rotated triangle antenna). However, at a lower rotational angle, as depicted in Figure 4b (30 degrees), the edge fails to support the excited modes since $l_{u1}$ does not intersect with the respective IFC, as illustrated in Figure 4i. Therefore, we have demonstrated various methods for antenna design and the control of selective mode excitation, leveraging curved edges, particularly through the technique of edge-assisted mode selection.

**Highly enhanced propagation length of g-HPs through selective mode excitation.**

The propagation length of phonon polaritons is an important parameter that quantifies the distance over which these hybrid light-matter quasiparticles can propagate before they are absorbed or scattered. A high propagation length is desirable in phonon polaritons for several reasons, such as increased sensitivity, enhanced energy transfer rate, longer range interactions, leading to improved device performance[32,49]. The modulation of phonon polaritons have been achieved through different techniques: modifying material properties, adjusting the strength of light-phonon coupling, reducing scattering and absorption through material selection and engineering techniques[50-53], and incorporating phonon resonators in the polaritonic structure[25].

In this study, we have modified the shape of the antenna to modulate the propagation length of the g-HP. Based on our theoretical analysis we identified that the propagation length of the polariton can extend beyond 50 µm for certain polaritonic modes (Supporting Information Note 5). Based on our experimental results, the newly designed triangular shape antenna shows longer propagation lengths. This is because the antenna is capable of selectively exciting polariton modes that can propagate an extended distance. Therefore, we can achieve the highest possible propagation length for calcite just by engineering the antenna shape to selectively excite the polariton mode. Previous research works have shown that most phonon propagation lengths are limited to a few microns[1,2,11,16,54], while the recently discovered g-HPs showed larger propagation lengths up to 20 microns[18]. In our experimental results we observe much longer propagation lengths exceeding 35 µm (Figure 5 c, e, g) for the triangular shape antenna. We can

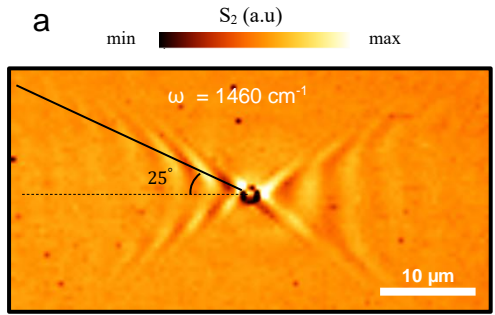

**a** S₂ (a.u)

ω = 1460 cm⁻¹

25°

10 μm

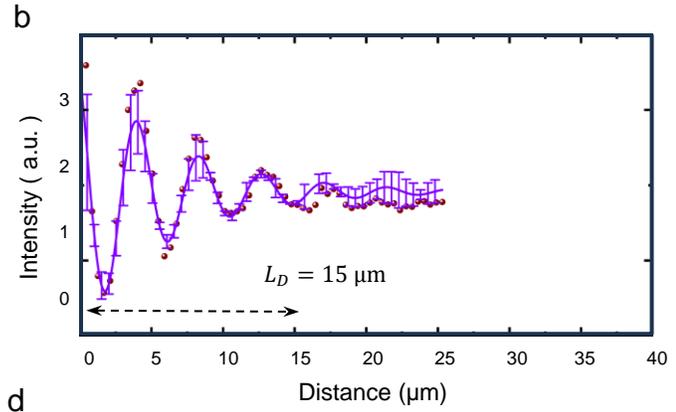

**b**

$L_D = 15\ \mu m$

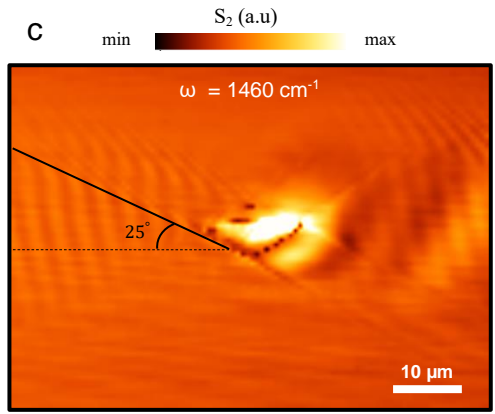

**c** S₂ (a.u)

ω = 1460 cm⁻¹

25°

10 μm

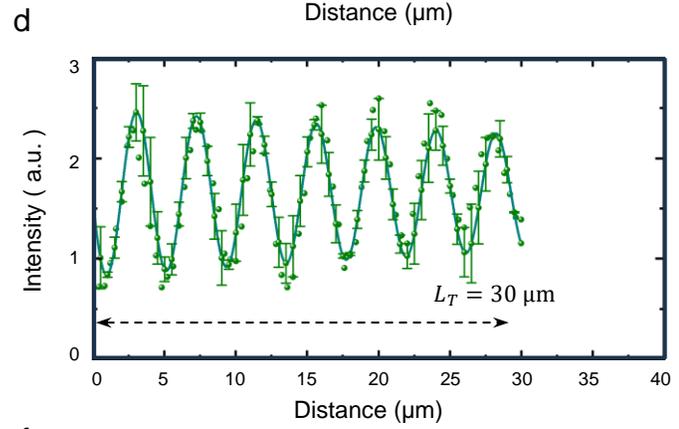

**d**

$L_T = 30\ \mu m$

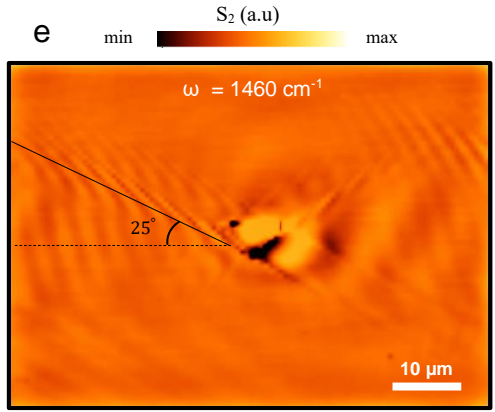

**e** S₂ (a.u)

ω = 1460 cm⁻¹

25°

10 μm

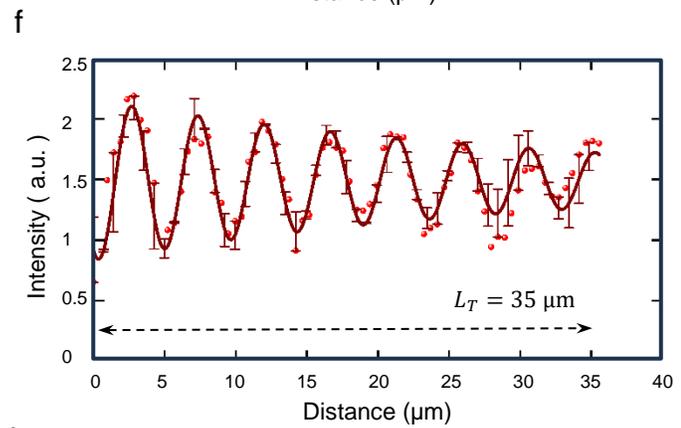

**f**

$L_T = 35\ \mu m$

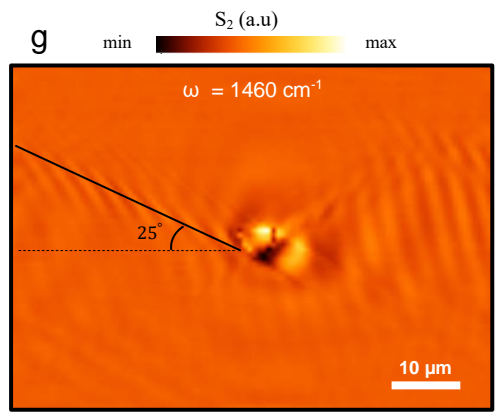

**g** S₂ (a.u)

ω = 1460 cm⁻¹

25°

10 μm

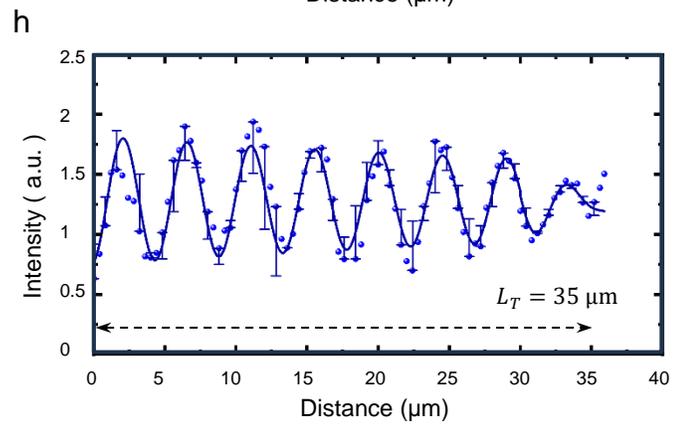

**h**

$L_T = 35\ \mu m$

**Figure 5. Highly enhanced propagation length of g-HPs through different shaped micro/nano antennas.** (a) Experimental near-field result of a disk-shaped nano antenna with a larger scanning area of $50 \times 25$ µm. (b) Line cut profile data for the disk-shaped nano antennas with an angle of $25°$ w.r.t the x axis. dark red balls represent the actual line cut profile data and the purple line is the sine damp fitted function explained in supplementary note 5 with a decay constant of $1.579 \times 10^5 \, m^{-1}$ . (c) Experimental near-field result of a triangular-shaped micro antenna with an inner angle $\delta = 33°$ and a scanning area of $70 \times 50$ µm. (d) Line cut profile with an angle of $25°$ w.r.t the x axis for the triangle-shaped micro antenna in c. green balls represent the actual line cut profile data and the dark green line is the sine damp fitted function (equation 21) with a decay constant of $1.22 \times 10^4 \, m^{-1}$ . (e) Experimental near-field result of a triangular-shaped micro antenna with an inner angle $\delta \approx 43°$ and a scanning area of $70 \times 50$ µm. (f) Line cut profile with an angle of $25°$ w.r.t the x axis for the triangle-shaped micro antenna in e. red balls represent the actual line cut profile data and the Brown is the sine damp fitted function (equation 21) with a decay constant of $3.0341 \times 10^4 \, m^{-1}$. (g) Experimental near-field result of a triangular-shaped micro antenna with an inner angle $\delta = 55°$ and a scanning area of $70 \times 50$ µm. (h) Line cut profile with an angle of $25°$ w.r.t the x axis for the triangle-shaped micro antenna in g. blue balls represent the actual line cut profile data and the dark blue line is the sine damp fitted function (equation 21) with a decay constant of $1.44 \times 10^4 \, m^{-1}$. The error bars in b, d, f and h represents the fitting uncertainty ( fitted value – actual value).

clearly see that the intensity of the polariton wave (Figure 5 d, f, h) obtained from the line cuts shown in respective SNOM images (Figure 5 c, e, g) for the triangular shape is strong even at 30-

35 μm, demonstrating the potential of having extended longer polariton wavefronts.

Both the shape and the size of the antenna have a huge effect on the propagation length of the polariton, as observed through the experimental and simulation results. The propagation length for the disk (Figure 5a) matches with previous reports[18]. The propagation length of the disk antenna according to the line cut profile in Figure 5b with a decay constant of $1.782 \times 10^5$ is around 15μm. The newly designed rectangle and triangle shaped antennas exhibit much stronger and longer propagations (Figure 1b, c). The triangular shape yields the longest propagation length, and the size and curvature of the antenna can further control the propagation length of the polariton. We have clearly observed extended propagation distances in the case of three triangular antennas, as depicted in Figure 5c,e and g. In Figure 5c, the triangular antenna features an inner angle ($\delta = 33°$) to entirely avoid edge assisted mode selection. Figure 5d provides a line cut profile (indicated by the black line in Figure 5c) of the polaritons selectively excited by the triangular antenna. Remarkably, even at the end of the line cut profile, the waveform retains substantial intensity (with a decay constant of $1.22 \times 10^4 \ m^{-1}$), indicating the potential for considerably longer propagation lengths. This remarkable outcome can be attributed to the fact that the triangular antenna has effectively excited long-propagating polariton modes. Moving to Figure 5e, the inner angle δ is set precisely at $\delta = 43° \pm 2°$, where the angle can almost have edge assisted mode selection at excitation wavenumber $\omega = 1460 \ cm^{-1}$. The line cut profile of this antenna (Figure 5f) also showcases high propagation length(with a decay constant of $3.0341 \times 10^4 \ m^{-1}$), but with some discrete mode excitations. In contrast, the triangular antenna with a larger inner angle ($\delta = 55°$) (Figure 5g) produces a polariton with edge assisted mode selection. This scenario showcases remarkably high intensity even at the end of the line profile with a decay constant of $1.44 \times 10^4 \ m^{-1}$, highlighting the potential for achieving greater propagation distances. In this case we can obtain single mode parallel fringes (parallel to the AC edge) and we can maintain the propagation to larger distances.

**Conclusion**

In conclusion, our experimental work has demonstrated the remarkable selective mode excitation of g-HPs using an asymmetric launching antenna on a calcite crystal surface. By manipulating the antenna's shape, orientation, and excitation wavelength we have shown that the polariton modes can be selectively excited to achieve high directionality and high propagation length of g-HPs. The mode selection can be flexibly controlled by edge assisted selective mode excitation technique that we demonstrate in the manuscript. Moreover, with the help of s-SNOM system, our near-field imaging experiments have revealed that the g-HP excited by the triangular antenna can propagate over a very long distance (more than 35 μm). All experimental results are consistent compared with simulation results.

Our work not only advances our understanding of the fundamental physics underlying g-HPs, but also offers new opportunities for developing a wide range of nanophotonic applications. By taking advantage of the interplay between structural engineering and g-HP theory, we have proved that anisotropic materials such as calcite have the potential to serve as a good platform for developing next-generation nanophotonic devices. In particular, such highly directional, long-propagating g-HP modes could be valuable for quantum information processing, on-chip optical communication, and high-precision sensing applications. The technique introduced here is general, and can be readily extended to other anisotropic materials, as well as other types of polariton modes. We anticipate that our work will stimulate further research in this rapidly growing field.

**Methods**

**Numerical Simulations**

We use COMSOL Multiphysics 5.1 software to simulate the electric field distribution of ghost polaritons emitted by gold antenna. Under oblique incidence (p-polarized plane wave, the incident angle is set as $\varphi = 60°$), the gold antenna will launch polariton wave propagating along the surface of calcite sample. After calculating the scattered electric field using finite element method, the intensity of electric field (recorded at 100 nm above the calcite surface) is plotted and compared with experimental results.

As for the permittivity of calcite crystal, we used a Lorentz oscillator model*:

$$\varepsilon_\perp = \varepsilon_{\infty,1}\left(1 + \frac{\omega_{LO,1}^2 - \omega_{TO,1}^2}{\omega_{TO,1}^2 - \omega^2 - i\omega\Gamma_1} + \frac{\omega_{LO,2}^2 - \omega_{TO,2}^2}{\omega_{TO,2}^2 - \omega^2 - i\omega\Gamma_2}\right) \qquad (2)$$

$$\varepsilon_\parallel = \varepsilon_{\infty,3}\left(1 + \frac{\omega_{LO,3}^2 - \omega_{TO,3}^2}{\omega_{TO,3}^2 - \omega^2 - i\omega\Gamma_3}\right) \qquad (3)$$

The optic axis is initialized to be aligned with y-axis and is then rotated with respect to x-axis by $\theta = 23.3°$.

The parameter of gold antenna is given as follows: for the "disk" antenna, its radius is 0.75 μm; for the "rectangle" antenna, its length is 3.2 μm, while its width is 1.4 μm; for the "triangle" antenna, its length is 8.4 μm, while its width is 4.9 μm. For all simulations, the thickness of gold antenna is set to be 50 nm. More simulation details can be found in Supplementary Information (Supporting Information Note 1 and 2).

## Materials and Fabrication

We used a commercially available polished calcite substrate (size: 10mm×10mm×0.5mm) that was prepared by mechanical cleavage from bulk calcite single crystal (trigonal structure). Electron beam lithography was used to fabricate the gold micro/nano antennas on the calcite substrate. The patterns were written on the resist (ARP: 6200/0.9) spin coated on the substrate (at 4000rpm for 1 minute and baked at 150°C for 1 minute). Cr (5 nm) and Au (45 nm) were deposited using e-beam evaporation and the standard lift off procedure (ZEP remover overnight) was used to finalize the fabrication.

## s-SNOM measurements

A commercially available s-SNOM (from Neaspec) that is based on a tapping mode atomic force microscope (AFM) was used to perform the real-space imaging of the polaritonic patterns. The tip was controlled to maintain a tapping amplitude of 70 nm to make sure that the tip can safely scan with the presence of the 50nm high micro/nano antennas. The Pt coated AFM tip is illuminated using a quantum cascade laser (QCL) at an angle about $60°$ with respect to the substrate surface. The QCL wavelength can be tuned from 1310 $cm^{-1}$ to 1705 $cm^{-1}$.

## Data Availability

The data and the code that supports the findings of this research are available from the corresponding authors upon reasonable request.

## Corresponding Author

* To whom correspondence should be addressed: Yuerui Lu (yuerui.lu@anu.edu.au), Zongfu Yu (zyu54@wisc.edu)

**Author Contributions**

Y.L conceived the project. M.P.S designed and fabricated the samples supervised by Y.L. M.P.S carried out the near-field imaging experiments with the help of H.Q. Q.Z formed the analytical calculations and carried out the numerical simulations supervised by Z.Y. M.P.S analyzed the data with the help of Q.Z, H.Q and X.S. M.P.S and Q.Z drafted the initial manuscript with the help of X.S, Z.L and it was supervised by Y.L, Z.Y and S.A.M.

**ACKNOWLEDGMENT**

The authors acknowledge funding support from ANU PhD student scholarship, Australian Research Council (ARC; numbers DP220102219, DP180103238, LE200100032) and ARC Centre of Excellence in Quantum Computation and Communication Technology (CE170100012). Qingyi Zhou and Zongfu Yu acknowledge support from National Science Foundation's QLCI-CI: Hybrid Quantum Architectures and Networks. . S.A.M. acknowledges the Lee-Lucas Chair in Physics. We would also like to thank Prof. Guangwei Hu for helpful discussions regarding numerical simulation.

Supporting Information

# Long-Propagating Ghost Phonon Polaritons Enabled by Selective Mode Excitation.

*Manuka P. Suriyage [1#], Qingyi Zhou[2#], Hao Qin[1], Xueqian Sun[1], Zhuoyuan Lu[1], Stefan A. Maier[4], Zongfu Yu[2*] and Yuerui Lu[1,3*]*

[1]School of Engineering, College of Engineering, Computing & Cybernetics, the Australian National University, Canberra, ACT, 2601, Australia

[2]Department of Electrical and Computer Engineering, University of Wisconsin-Madison, Madison, WI 53706, USA

[3]ARC Centre of Excellence in Quantum Computation and Communication Technology ANU node, Canberra, ACT 2601, Australia

[4]School of Physics and Astronomy, Monash University Clayton Campus, Melbourne, Victoria 3800, Australia

# Those two authors contribute equally to this work.

\* To whom correspondence should be addressed: Yuerui Lu (yuerui.lu@anu.edu.au), Zongfu Yu (zyu54@wisc.edu)

## Supporting Information Note 1: Theory of g-HP mode in rotated calcite

In this section, by solving Maxwell's equations in frequency domain[1], we verify the existence of ghost polariton mode at the interface between calcite and vacuum.

Consider the interface between anisotropic calcite crystal ($z < 0$) and vacuum ($z \geq 0$). The optic axis (OA) of calcite crystal is initialized to be aligned with y-axis. Then the OA is rotated by an angle θ with respect to x-axis. Therefore, the permittivity tensor of calcite substrate can be written as:

$$\bar{\bar{\varepsilon}} = R_x(-\theta) \cdot \begin{pmatrix} \varepsilon_\perp & 0 & 0 \\ 0 & \varepsilon_\parallel & 0 \\ 0 & 0 & \varepsilon_\perp \end{pmatrix} \cdot R_x(\theta), \tag{S1}$$

where $R_x(\theta)$ represents the rotation matrix with respect to x-axis. The permittivity tensor can be derived as:

$$\bar{\bar{\varepsilon}} = \begin{pmatrix} \varepsilon_\perp & 0 & 0 \\ 0 & \varepsilon_\parallel \cos^2\theta + \varepsilon_\perp \sin^2\theta & (\varepsilon_\parallel - \varepsilon_\perp)\sin\theta\cos\theta \\ 0 & (\varepsilon_\parallel - \varepsilon_\perp)\sin\theta\cos\theta & \varepsilon_\perp \cos^2\theta + \varepsilon_\parallel \sin^2\theta \end{pmatrix}. \tag{S2}$$

Now we are ready to start with the source-free Maxwell's equation. Note that only electric field $\vec{E}$ is kept:

$$\nabla \times \nabla \times \vec{E} - k_0^2 \bar{\bar{\varepsilon}} \cdot \vec{E} = 0, \tag{S3}$$

where $k_0$ denotes the wave vector inside vacuum. We'd like to focus on eigenmodes with $e^{i\vec{k}\cdot\vec{r}}$ dependence. The above Maxwell's equation is thus simplified:

$$\vec{k} \times \vec{k} \times \vec{E} + k_0^2 \bar{\bar{\varepsilon}} \cdot \vec{E} = 0. \tag{S4}$$

Denote the components of $\vec{k}$ as $\vec{k} = [k_x, k_y, k_{2z}]$. The fact that this equation has non-zero solution for $\vec{E}$ indicates that the following determinant equals zero:

$$\begin{vmatrix} -k_{2z}^2 - k_y^2 + \varepsilon_\perp k_0^2 & k_x k_y & k_{2z} k_x \\ k_x k_y & -k_{2z}^2 - k_x^2 + k_0^2(\varepsilon_\parallel \cos^2\theta + \varepsilon_\perp \sin^2\theta) & k_{2z} k_y + k_0^2(\varepsilon_\parallel - \varepsilon_\perp)\sin\theta\cos\theta \\ k_{2z} k_x & k_{2z} k_y + k_0^2(\varepsilon_\parallel - \varepsilon_\perp)\sin\theta\cos\theta & -k_x^2 - k_y^2 + k_0^2(\varepsilon_\perp \cos^2\theta + \varepsilon_\parallel \sin^2\theta) \end{vmatrix}. \tag{S5}$$

With the help of Wolfram Mathematica 12.3, we conduct a factorization, which leads to two equations:

$$k_x^2 + k_y^2 + k_z^2 = \varepsilon_\perp k_0^2, \tag{S6}$$

$$\frac{(k_y \cos\theta + k_{2z}\sin\theta)^2}{\varepsilon_\perp} + \frac{k_x^2 + (k_y\sin\theta - k_{2z}\cos\theta)^2}{\varepsilon_\parallel} = k_0^2. \tag{S7}$$

The first equation corresponds to ordinary wave (o-wave), while the second equation corresponds to extraordinary wave (e-wave).

For given $k_x$ and $k_y$, the value of $k_{2z}$ can be derived by solving the quadratic equation:

$$k_{2z} = \frac{-k_y(\varepsilon_\parallel - \varepsilon_\perp)\sin\theta\cos\theta \pm \sqrt{\Delta}}{\varepsilon_\perp \cos^2\theta + \varepsilon_\parallel \sin^2\theta}, \tag{S8}$$

where $\Delta = \varepsilon_\perp[\varepsilon_\parallel k_y^2 + (\varepsilon_\perp \cos^2\theta + \varepsilon_\parallel \sin^2\theta)(k_x^2 - \varepsilon_\parallel k_0^2)]$. When $\Delta < 0$, $k_{2z}$ becomes complex, leading to the g-HP modes we want.

**Supporting Information Note 2: Detailed information of numerical simulations**

In this section, we introduce more details regarding the numerical simulations presented in the main text. We use COMSOL Multiphysics 5.1 software for all the numerical simulations. As we've mentioned in the main text, we use a Lorentz oscillator model1 to calculate the permittivity of calcite substrate (Figure S1):

$$\varepsilon_\perp = \varepsilon_{\infty,1}\left(1 + \frac{\omega_{LO,1}^2 - \omega_{TO,1}^2}{\omega_{TO,1}^2 - \omega^2 - i\omega\Gamma_1} + \frac{\omega_{LO,2}^2 - \omega_{TO,2}^2}{\omega_{TO,2}^2 - \omega^2 - i\omega\Gamma_2}\right) \tag{S9}$$

$$\varepsilon_\parallel = \varepsilon_{\infty,3}\left(1 + \frac{\omega_{LO,3}^2 - \omega_{TO,3}^2}{\omega_{TO,3}^2 - \omega^2 - i\omega\Gamma_3}\right) \tag{S10}$$

The OA is initialized to be aligned with y-axis and is then rotated with respect to x-axis by $\theta = 23.3°$.

For Au, permittivity data measured experimentally is used[2]. Inside the wavelength range we're interested in ; Au has a negative permittivity whose absolute value is very large. Thus, the gold antenna performs very similar to a perfect electric conductor (PEC): the electric field inside antenna remains very small.

It has been found that when solving Maxwell's equations in frequency domain, anisotropic material such as calcite may lead to poor convergence if iterative solver is used. Therefore, the direct solver "PARDISO" is used in all our COMSOL simulations.

Simulation details for Figure 1

For the simulations shown in Figure 1, the simulation domain size in xy-plane is 75 $\mu m$ × 50 $\mu m$. The frequency is set as $\omega = 1460 \ cm^{-1}$. The gold antenna, with a thickness of 50 nm, is put at the center of xy-plane, just above the vacuum-calcite interface.

With the help of total-field scattered-field (TF-SF) technique, a p-polarized plane wave serves as the incident wave (incident angle set to be $60°$). By putting a monitor at 100 nm above the calcite's surface, the near-field distribution can be recorded. Finally, the corresponding light intensity $\left|\vec{E}\right|^2$ is plotted, as shown in Figure 1d-f. Note that only the middle part of the simulation domain is shown in Figure 1, while artifacts close to the domain boundaries are cropped out. The corresponding Fourier transform (FT) results are obtained with the help of MATLAB R2021b's "fft2()" function.

Here we provide the shape parameters of the three gold antennas we've used:

(1) For the "disk" antenna, its radius is 0.75 μm.

(2) For the "rectangle" antenna, its length is 3.2 μm, while its width is 1.4 μm. Instead of using a perfect rectangular box, in COMSOL we construct this antenna using one rectangular box and two cylinders (with radius 0.7 μm). Therefore, we can keep the antenna shape used in simulation consistent with the one used in experiment.

(3) For the "triangle" antenna, its length is 8.4 μm, while its width is 4.9 μm. Based on similar reason, in COMSOL this antenna consists of three cylinders, three rectangular boxes, as well as one prism (shown in the inset of Figure S12a).

Simulation details for Figure 2

For the simulations shown in Figure 2, the simulation domain size in xy-plane is 75 μm ×50 μm. Similarly, a gold antenna, with a thickness of 50 nm, is put above the vacuum-calcite interface. Note that here to show the directionality of emitted polariton, only the "triangle" antenna is used.

In order to demonstrate that the directionality of emitted polariton wave strongly depends on the frequency, the light intensity distributions for multiple different frequencies ($\omega = 1430\ cm^{-1}$, $1450\ cm^{-1}$, $1470\ cm^{-1}$) are plotted.

Parameters regarding the incident wave, as well as other hyper-parameters, remain the same compared with Figure 1.

Simulation details for Figure 4

For the simulations shown in Figure 4, the simulation domain size in xy-plane is 75 μm ×50 μm. The frequency is set as $\omega = 1460\ cm^{-1}$. Similar to Figure 2, only the "triangle" antenna is used. The only difference is that here we try to rotate the antenna.

In order to demonstrate that the directionality of emitted polariton wave depends on the rotation angle of gold antenna, the light intensity distributions for multiple rotation angles ($\sigma = 0°, 30°$, $60°, 90°$) are plotted.

Other parameters remain the same compared with Figure 2.

**Supporting Information Note 3: Selective mode excitation of different shape nano antennas: directionality measurements and excited mode identification.**

There are four polaritonic branches that are being studied in this work. The four branches have been defined based on the direction of polariton mode's group velocity. More specifically, we define the "upper" branch and "lower" branch based on the y component of group velocity (energy going to +y direction gives U, while -y gives L). On the other hand, the "+" and "-" are defined based on the x component of group velocity (energy going to +x direction gives +, while -x gives -). Combining these two would give 4 branches, which we call U+, U-, L+, L-. These 4 branches correspond to the 4 quadrants of the reciprocal space:

$$U+: k_x > 0, k_y < 0;\ U-: k_x < 0, k_y < 0.$$

$$L+: k_x > 0, k_y > 0;\ L-: k_x < 0, k_y > 0.$$

The two different angle parameters shown in Figure S2b and c are the angles of the incident wave φ (w.r.t the z axis) and $\emptyset$ (within x-y plane) that can be used to create an asymmetric distribution in polaritonic branches. This has been investigated recently by a few studies where the broken of symmetry for g-HPs[3], hyperbolic PhPs[4] and even bloch modes[5]. These techniques do have the capability to get broken symmetry but limited in controlling the directionality of the polariton. Even when the incident angle of the excitation wave changes the formation of the asymmetric polaritonic rays depends on the weak and strong local fields generated on the left and right side of the nano antenna[1]. Therefore, the left (-) and the right (+) polaritonic branches are no longer symmetric. The strong local fields (larger local photonic density of state) on the (+) polaritonic branch results in the superposition of many polaritonic modes with different wavevectors and thus showcases a ray like propagation. Figure S2d-f represents the symmetric and asymmetric distribution of the polaritons resulted due to the change of incident angle. It is clear that the normal incident (Figure S2d) showcases a symmetric polariton distribution with similar propagations in all four polaritonic branches and this can be clearly visualized in the momentum space as illustrated in Figure S2g. There are four bright regions in the momentum space representing the four k vectors ($k_{L-}, k_{L+}, k_{U-}, k_{U+}$)) with an equal magnitude that justifies the symmetric distribution in the polariton wavefronts.

When φ changes from $90°$ to $60°$ as illustrated in Figure S2e the U- and L+ branches are much more intense and this intensity of those two branches increases if the incident angle is further reduced as evident in Figure S2f (where $\varphi = 30°$). The corresponding FFT results (Figure S2h,i) confirm the asymmetry in the propagation through the brightness of the modes excited for each branch. Apart from this the asymmetry of the polariton can be achieved through changing the polarization angle ∅ (Figure S2c) as recently demonstrated in this study[3]. Even in this study the symmetry is broken in a way that is achieved through changing of φ but with that only the intensity of the U- and the L+ branches can be controlled.

Here in our research, we have suppressed the lower branch completely by just changing the shape and the size of the micro/nano antenna and to demonstrate this we used two different techniques to evaluate the directionality of the polariton as explained in the main text. First the directionality of the generated g-HP is determined by comparing the polariton intensities of the upper and lower branches. The directionality constant β is defined as in Equation 1 in the main text. For the polaritons generated by the three shapes we calculated the amplitude intensity of the upper and lower branches of the polariton by considering a line cut profile for the obtained results (Figure S3a, d and g). The line cuts were produced such that the polariton intensity contrast (ratio between the highest and the lowest values) is the maximum. For the three main nano antenna shapes the directionality constant results are as follows (amplitude values shown in Figure S3c, f and i).

$$\beta_{disk} = \left| \frac{0.113}{0.108} \right| = 1.046$$

$$\beta_{rectangle} = \left| \frac{0.446}{0.149} \right| = 2.99$$

$$\beta_{triangle} = \left| \frac{0.474}{0.109} \right| = 4.348$$

Therefore, we can clearly see that the directionality can be modulated by changing the shape of the antenna.

$$\beta_{triangle} > \beta_{rectangle} > \beta_{disk} \tag{S11}$$

Similarly for different orientations of the triangle (Figure S4) w.r.t the calcite surface we can obtain different directionality constant values, confirming that the directionality can be modified just by changing the orientation of the nano antenna (further explained in S4). The following β values are calculated based on the amplitude intensities of the upper and lower polariton line cuts of the different orientations of the triangular nano antenna σ= 0˚, 30˚, 60˚, 90˚ (Figure S4e-h).

$$\beta_0 = \left| \frac{0.474}{0.109} \right| = 4.348$$

$$\beta_{30} = \left| \frac{0.251}{0.062} \right| = 4.04$$

$$\beta_{60} = \left| \frac{0.427}{0.233} \right| = 1.83$$

$$\beta_{90} = \left| \frac{0.308}{0.179} \right| = 1.72$$

$$\beta_0 > \beta_{30} > \beta_{60} > \beta_{90} \tag{S13}$$

Secondly, simulation and experimental results were subjected to Fourier transform to reveal two distinct hyperbolic iso-frequency contours (IFCs) in k-space. All the simulation results are $|E_z|$ field distribution on xy plane, while all the experimental results shown are amplitude signal $S_2$ obtained through s-SNOM. The amplitude (A) and the phase (∅) of the polariton can be extracted from the SNOM second harmonic amplitude (O2A; Figure S5a) and phase (O2P; Figure S5b) images. This Amplitude and phase data is used to calculate the E field which is then used to produce the FT results as illustrated in Figure S5c.

$$E = A \times e^{i\emptyset} \tag{S12}$$

There are either single or multiple modes excited by the antenna depending on various parameters as evident by the FFT results (Figure S6). The k vectors and the group velocities drawn in main text Figure 1 were identified as illustrated in Figure S3j-o. Panels J to L represents the same set of FT results shown in main text Figure 1, but we have shown how the FFT amplitude intensities were measured using a line cut profile to determine the different modes excited by the differently shaped antennas. The line cut profiles (Figure S3m-o) showcase the excitation efficiency of the polariton modes by each shaped antenna. The dot and the rectangle are exciting some low k modes with a higher efficiency, but it is gradually decreasing showcasing low efficiencies in high k mode excitation. One of the two modes shown in Figure 1g (represented with wave vector $K_{L-}$) is the $M_{1D}$ mode. This is the largest k vector with a reasonable coupling efficiency that a 1.5 µm disk can excite. A smaller disk can excite much larger k vectors. The 1.5 µm roughly produces all the low k vectors in between $K_{L-}$ and $K_{u-}$. Similarly, the $K_{u-}$ mode shown in Figure 1h corresponds to the $M_{1R}$ mode which is the mode that the rectangle antenna excites with the highest efficiency. But the FT for the triangle (Figure S3l) and the line cut profile (Figure S3o) clearly shows that the triangle antenna can excite multiple discrete modes with different efficiencies. The $K_{u-}$ wave vector in Figure 1i is drawn to represent one of these highly efficient modes ($M_{2T}$) that a triangle antenna excite.

The comparison of the intensities of each branch through β and the polaritonic mode analysis through the FT results were crucial to properly identify the selective mode excitation that is achieved through engineering the micro/nano antenna. Even for the rotated triangles the excited modes are clearly visible through the FFT results shown in Figure S4i-l. The FFT results for Figure 1 from both experimental and simulation data, is illustrated in Figure S6. These IFCs correspond to the left and right parts of the polariton generated by each shape, with the outer and inner IFCs representing the left and right polaritons, respectively. It can be observed that the left polariton exhibits a concaved wavefront, whereas the right polariton wave is diffractionless.

We intended to design the micro/nano antenna to achieve more directionality and long propagation First, we performed a series of simulations (see Figure S7) to determine the effect of different diameters of a disk antenna and found that the propagation length increases with the increasing diameter. To confirm, we investigated polariton excitation behavior through the experimental near field results for a few different disk antennas with increasing diameters (see Figure S8a-c). When the diameter increases the electric field distribution gets complex allowing multiple curved points in the antenna edge to allow exciting polaritons. As illustrated in Figure S8c the disk antenna (radius = 2 µm) showcases four curved points (A-D) which polariton rays are excited. When the diameter decreases these points come closer to each other resulting in highly confined polariton excitations. If the disk structure is extended in the y or x axis to generate an ellipse shape, we observed that two curved points dominate in exciting polariton modes. As illustrated in Figure S8d the vertical ellipse with a major axis in the y direction (major axis = 8 µm, minor axis = 4 µm) excites two portions at curved points C and D. The interference fringes of both those excited polaritons (at points C and D) towards the left side (- branch) can be observed clearly. Then we extended the design to a triangle which consists of three curved points (vertices of the triangle) named A, B and C (see Figure S8i left panel; nearfield image). These three vertices will act as the polariton excitation points, and the curvature will determine the most efficient modes excited by the curved points of the triangle micro antenna.

For a triangle shaped antenna, the left side polariton branches (- branch) are assisted by the collective excitation by the two vertices A and C of the triangle. By changing the curvature of the curved point A and C, the excited mode can be altered. Then the inclined edge AC of the triangle can either enhance or suppress the excited polariton modes. This is explained in S4. Therefore, the selective excitation of a triangle is determined by the curvature of the three vertices and the edge orientation of the triangular shape.

**Supporting Information Note 4: Edge assisted mode selection.**

The rectangular and triangular-shaped micro antennas employed here for polariton excitation feature notably larger physical edges. These edges can direct reflected light towards the same direction that is perpendicular to the specific physical edge of the nano antenna under specific conditions[6]. At the same time the edge assists to enhance or suppress the excited polariton modes. This section provides insights into the design parameters governing these physical edges for the purpose of controlling edge assisted mode selection. The anisotropy of the calcite material, which is a hyperbolic material, can be effectively characterized through its IFC, represented in the following manner:

$$\frac{x^2}{c_x} - \frac{y^2}{c_y} = 1 \tag{S14}$$

Where, $c_x = Re(\varepsilon_\parallel)$ and $c_y = Re(\varepsilon_\parallel \sin^2\theta + \varepsilon_\perp \cos^2\theta)$

As illustrated in Figure 3a $\boldsymbol{k_i}$ represents the incident wave in the momentum space and the $\boldsymbol{k_i}$ vector can be written as:

$$\boldsymbol{k_i} = k_{ix} + k_{iy} \tag{S15}$$

Where $k_{ix} = \cos\varphi \times \cos\emptyset$ and $k_{iy} = \cos\varphi \times \sin\emptyset$.

In our work the tip generated polariton is not reflected by the physical edge of the antenna. The tip generated polariton does not propagate long distances in calcite since the tip is extremely small. As illustrated Figure 3a schematic if the vertices of the triangle antenna (A and B) excite a mode which has a k vector perpendicular to the AC physical edge the edge supports those modes and enhances the excitation efficiency to allow a single mode to propagate. Interestingly the curvature of the vertex C can be adjusted to allow polariton modes in the U- branch which has wave vectors perpendicular to the edge and then the edge can enhance its coupling efficiency. Therefore, the B curved vertex, AC physical edge and the C vertex collectively couples the energy of the incident light to a single polariton mode allowing parallel fringes to the physical edge.

We can get the relationship between the internal angle of the triangle δ and the selectively excited k vector $\boldsymbol{k_e}$ through the following:

$$\delta = 90° - \tan^{-1}\left(\sqrt{\frac{c_y}{c_x} \times \frac{(k_e^2 - c_x)}{(k_e^2 + c_y)}}\right) \tag{S16}$$

According to Equation S16, for a specific excitation frequency below a critical δ angle the edge cannot support any edge assisted modes. To verify this, we simulated two distinct triangles, representing the blue and green colored triangles from Figure 3a ( with internal angles $\delta_1$ and $\delta_3$). The simulations were performed with an illumination frequency of $\omega = 1460\ cm^{-1}$. The simulation outcome for the larger triangle ($\delta_1 = 70°$) is illustrated in Figure S9a. The FT result for this polariton excitation (Figure S9c) shows one bright spot at the coincident point of the IFC and line $l_{t1}$. This perfectly validates the edge assisted mode selection that can be achieved by varying the angle of the AC physical edge. In contrast, the simulated a triangle with $\delta_3 = 30°$ (Figure S9b) does not show parallel fringes to the AC physical edge, and the corresponding FT result reveals multiple excited modes. However, there is no intersection point between the IFC and line $l_{t3}$ confirming the absence of edge-supported mode selection.

## Supporting Information Note 5: Propagation length based on different shapes and sizes of the nano antenna.

In this section, we introduce details regarding how the propagation length of g-HP modes can be predicted theoretically[4]. To remain consistent with S1, we assume plane wave solutions with $e^{i\vec{k}\cdot\vec{r}}$ dependence. The g-HP mode is a combination of eigenmodes inside air and eigenmodes inside calcite. Denote the wave vector of plane wave solutions inside air as $\vec{k}_1 = [k_x, k_y, k_{1z}]$, where $k_{1z} = i\sqrt{k_x^2 + k_y^2 - k_0^2}$. The eigenmodes can be decomposed into two independent solutions:

$$\text{TM: } \vec{E}^{\text{TM}} \sim [-k_x k_{1z}, -k_y k_{1z}, k_x^2 + k_y^2];$$

$$\text{TE: } \vec{E}^{\text{TE}} \sim [-k_y, k_x, 0].$$

For eigenmodes inside calcite, denote the wave vector as $\vec{k}_2 = [k_x, k_y, k_{2z}]$. Again, two solutions exist: for the ordinary wave, we use $k_{2z} = -i\sqrt{k_x^2 + k_y^2 - \epsilon_\perp k_0^2}$ to make sure that the electric field decays to zero when $z \to -\infty$; for the extraordinary wave, two $k_{2z}$ solutions can be found by solving quadratic equation, as stated in the previous section. Note that only the solution which satisfies $\text{Im}[k_{2z}] < 0$ will be used. The electric field $\vec{E}^o$ (ordinary wave) and $\vec{E}^e$ (extraordinary wave) can be derived by solving the source-free Maxwell's equation:

$$\vec{k}_2 \times \vec{k}_2 \times \vec{E} + k_0^2 \bar{\bar{\epsilon}} \cdot \vec{E} = 0. \tag{S17}$$

In order to derive the dispersion relationship of g-HP modes, we now focus on a linear combination of the above 4 modes and impose the continuous conditions for tangential fields $E_x$, $E_y$, $H_x$, $H_y$ at $z = 0$. The corresponding 4 linear equations are.

$$\begin{pmatrix} E_x^o & E_x^e & -E_x^{\mathrm{TM}} & -E_x^{\mathrm{TE}} \\ E_y^o & E_y^e & -E_y^{\mathrm{TM}} & -E_y^{\mathrm{TE}} \\ H_x^o & H_x^e & -H_x^{\mathrm{TM}} & -H_x^{\mathrm{TE}} \\ H_y^o & H_y^e & -H_y^{\mathrm{TM}} & -H_y^{\mathrm{TE}} \end{pmatrix} \begin{pmatrix} a^o \\ a^e \\ a^{\mathrm{TM}} \\ a^{\mathrm{TE}} \end{pmatrix} = \vec{0} \tag{S18}$$

where $a^o/a^e$ and $a^{\mathrm{TM}}/a^{\mathrm{TE}}$ are the coefficients of eigenmodes inside air and calcite, respectively. The fact that this equation has non-zero solution for $\vec{a}$ indicates that the following determinant equals zero:

$$\begin{vmatrix} E_x^o & E_x^e & -E_x^{\mathrm{TM}} & -E_x^{\mathrm{TE}} \\ E_y^o & E_y^e & -E_y^{\mathrm{TM}} & -E_y^{\mathrm{TE}} \\ H_x^o & H_x^e & -H_x^{\mathrm{TM}} & -H_x^{\mathrm{TE}} \\ H_y^o & H_y^e & -H_y^{\mathrm{TM}} & -H_y^{\mathrm{TE}} \end{vmatrix} = 0 \tag{S19}$$

In order to find the isofrequency contour of g-HP modes, for any given $k_y$ the above equation is solved to find the corresponding $k_x$. We also impose the constraint that the imaginary part of in-plane wave vector $[\mathrm{Im}(k_x), \mathrm{Im}(k_y)]$ should be parallel to the group velocity $\vec{v}_g$. Combining the above conditions, both the real and imaginary part of $k_x$, $k_y$ can be calculated numerically with the help of Wolfram Mathematica 12.3.

The propagation length $L$ (Equation S20), defined as the length at which the electric field's magnitude decays to $1/e$, can therefore be derived as

$$L = \frac{1}{\sqrt{\mathrm{Im}(k_x)^2 + \mathrm{Im}(k_y)^2}} \tag{S20}$$

Here the relationship between the propagation length $L$ and the propagation direction has been plotted in Figure S10. The propagation direction is quantified by angle $\gamma$, which is the angle between the propagation direction and the $x$-axis. There are several points that worth paying attention to:

(1) The directionality angle increases when the frequency increases, which is consistent to literatures as well as our experimental results.

(2) The propagation length $L$ becomes larger for smaller angle $\gamma$. However, for very small $\gamma$ (corresponds to small $|k_y|$), no solution can be found numerically, meaning that g-HP mode does not exist if the angle is too small.

(3) The propagation length $L$ can be larger than 50 $\mu m$ at 1460 cm$^{-1}$ and 1470 cm$^{-1}$. This is consistent with the fact that we have observed a long-propagating g-HP mode ($L > 35 \mu m$) in our experiment.

So, propagation length certainly depends on both the size and the shape of the nano antenna since those specially designed nano antennas can selectively excite certain polaritonic modes. We further analyzed the effect of the shape towards the propagation length. By varying the radius of curvature (Figure S11) of the triangular shape, while maintaining a constant height, we observed that the propagation length decreases with increasing radius of curvature (Figure S12a). For the radii of curvature r =0.1 µm to r =2.1 µm we can only see a slight change in the propagation length around 20 µm. But when the radius of curvature changes to r = 2.5 µm, we clearly observe a decrease in the propagation length to 15 µm. Therefore, even with the same area of the orthographic projection of the shape we can see different propagation lengths based on the shape of the antenna. Although there isn't a huge difference when we compare the simulation results, a huge difference in the propagation length can be observed when comparing the experimental results. This is because the fabricated nano antennas are able to excite few modes extra that have a larger propagation length because of the complexities at the edges of the fabricated nano antenna. Similarly, the size of the disk antenna affects the propagation length and the polariton behavior as in Figure S7. The line cut profiles of those simulation results are illustrated in Figure S12b which shows that for a disk-shaped antenna, the propagation length decreases when the size of the antenna decreases, and the polaritons become much more diffraction less.

The line cut profile data taken from the four s-SNOM images in main text Figure 4a,c and g are fitted using a sine damp function as in Equation S21:

$$y(x) = y_0 + A \times e^{-dx} \times \frac{\sin(\pi(x - x_c))}{w} \qquad (S21)$$

Where; $y_0$ = offset, A = amplitude, d = decay constant, $x_c$ = phase shift, w = period

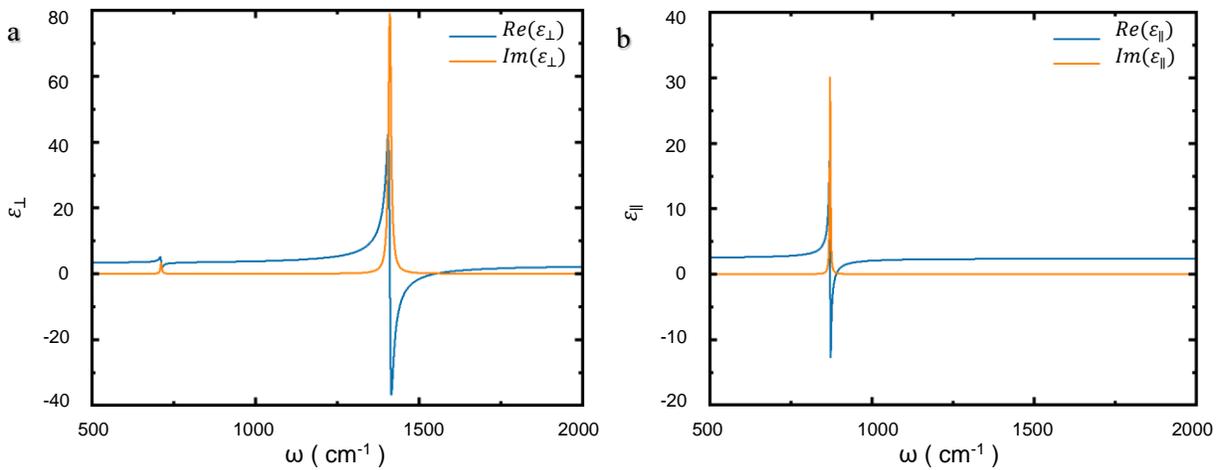

**Figure S1. Calcite's permittivity calculated using the Lorentz oscillator model.** (**a**) Real and imaginary parts of calcite's permittivity tensor component $\epsilon_\perp$. (**b**) Real and imaginary parts of calcite's permittivity tensor component $\epsilon_\parallel$.

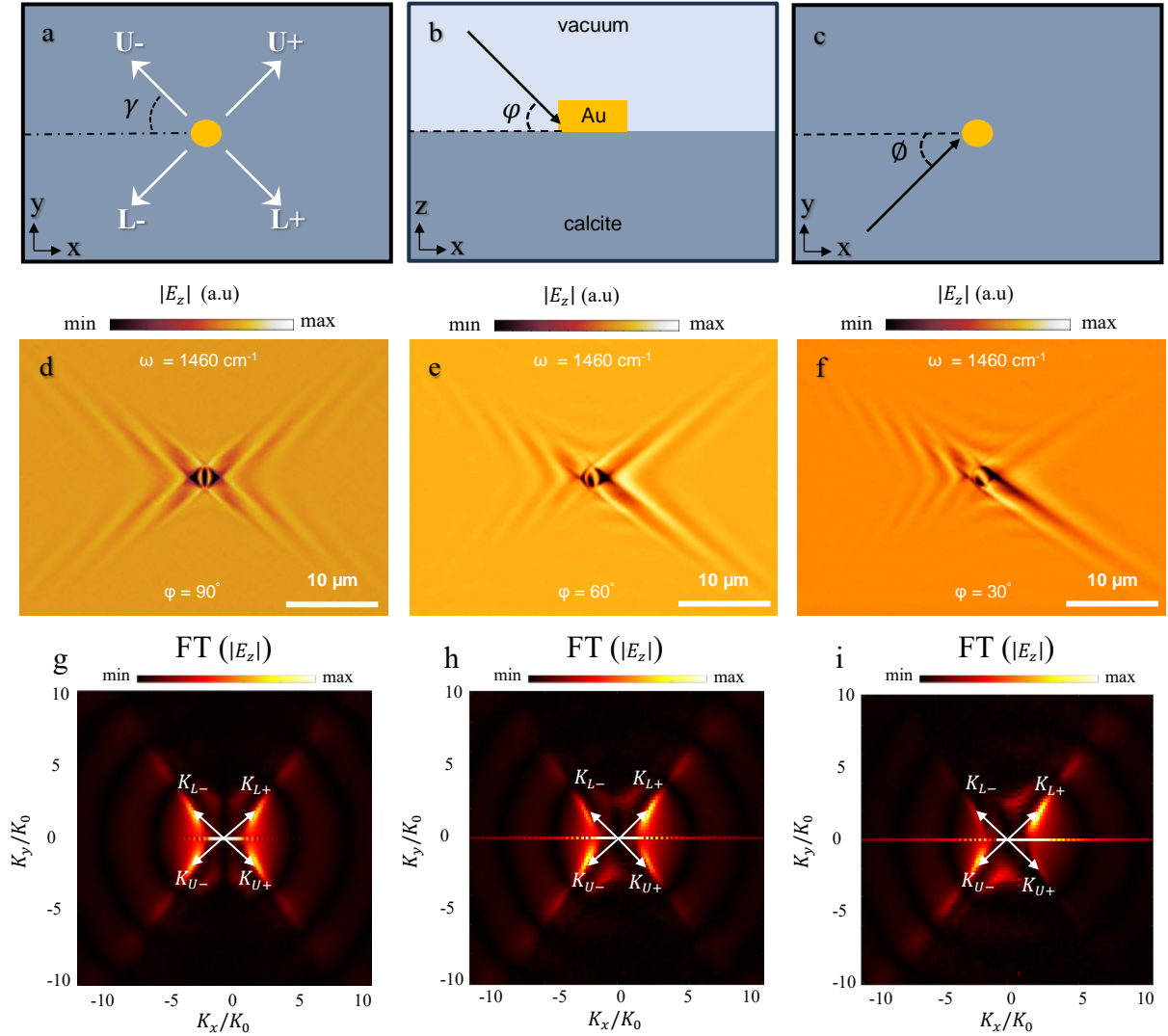

**Figure S2. Polariton parameters and introducing asymmetric excitation. (a)** Schematic illustration of the four branches of polaritons U+ (upper positive), U- (upper negative), L+ (lower positive), L- (lower negative) and the propagation direction angle γ. **(b)** Schematic illustration of the incident angle w.r.t z axis **φ**. (c) Schematic illustration of the polarization angle which is the incident angle of the photon within the x-y plane. **(d-f)**, Simulated near-field images of antenna-launched g-HPs at the illumination frequency of $\omega = 1460\ cm^{-1}$ with an incident angle of $\varphi = 90°, 60°, 30°$. **(g-i)**, Fourier transformed $|E_z|$, generated by the disk antennas in d-f. white arrows represent the wave vectors of the excited modes by the antenna.

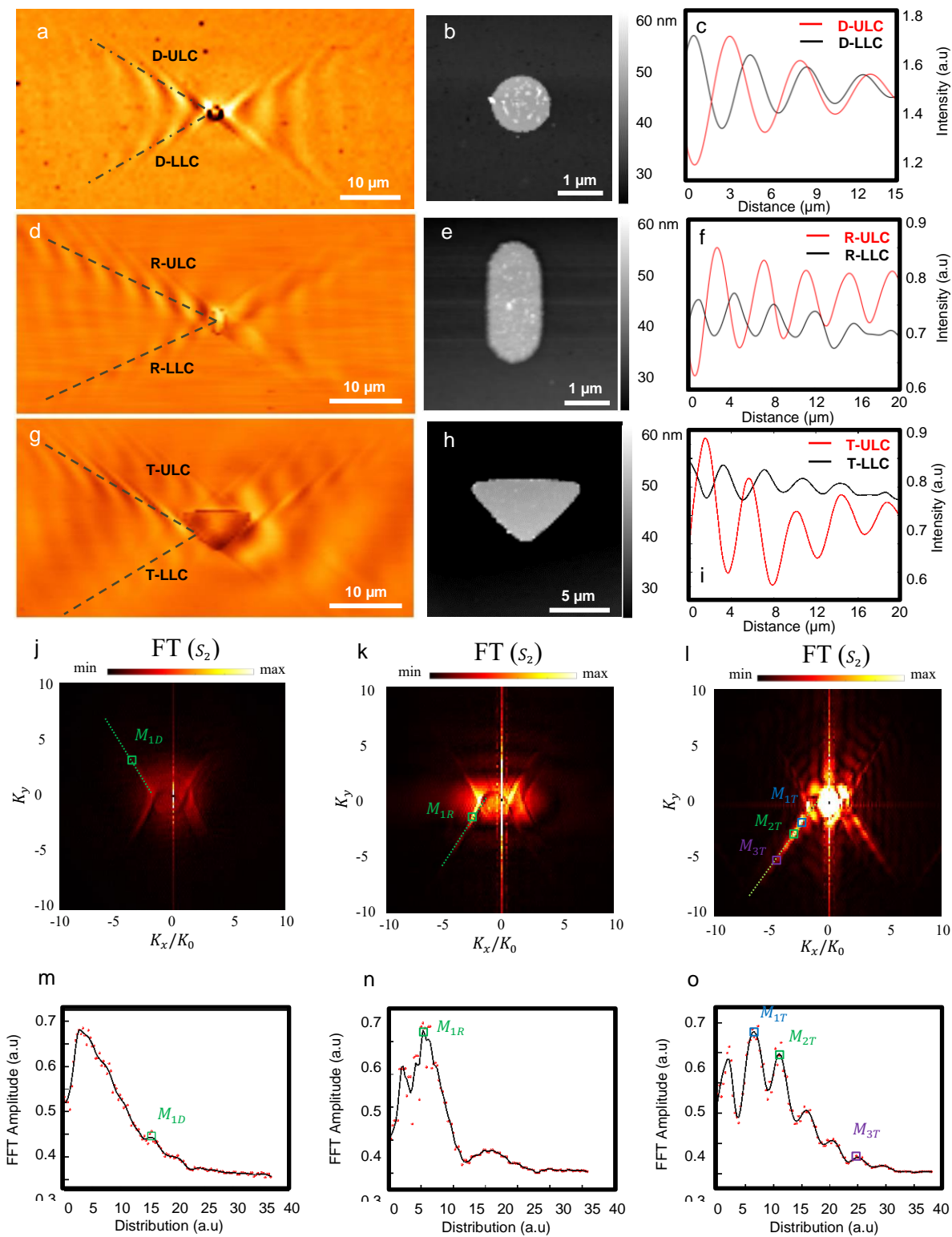

**Figure S3. Determination of directionality of g-HPs for different shapes.** (**a**) 50µm × 25µm
s-SNOM scan of a disk-shaped antenna marked with the upper line cut (D-ULC) and lower line cut (D-LLC). (**b**) AFM image of the disk antenna (diameter = 1.5 µm). (**c**) Upper and lower line cut intensity profiles of the disk antenna shown in A. (**d**) 50µm × 25µm s-SNOM scan of a rectangle shaped antenna marked with the upper line cut (R-ULC) and lower line cut (R-LLC). (**e**) AFM image of the rectangle antenna (W × H = 1.2 µm x 3.2 µm). (**f**) Upper and lower line cut intensity profiles of the rectangle antenna in D. (**g**) 50µm × 25µm s-SNOM scan of a triangle shaped antenna marked with the upper line cut (T-ULC) and lower line cut (T-LLC). (**h**) AFM image of the triangle antenna (L × H = 8.8 µm × 4.9 µm). (**i**) T-ULC and T-LLC intensity profiles of G. (**j-l**) Absolute value of the Fourier transform of the images shown in A, D, G. (**m-o**) FFT amplitude distribution along the line cut profiles shown in J, K and L. $M_{1D}$ represents a mode excited by the disk antenna related to the peak shown in panel M. $M_{1R}$ is the most prominent mode excited by the rectangular antenna corresponding to the peak shown in N. $M_{1T}$, $M_{2T}$, $M_{3T}$ are 3 discrete modes excited by the triangle shape antenna with a high efficiency corresponding to the peaks shown in o. Green dotted line is used to produce the intensity line cut profiles shown in panels m-o.

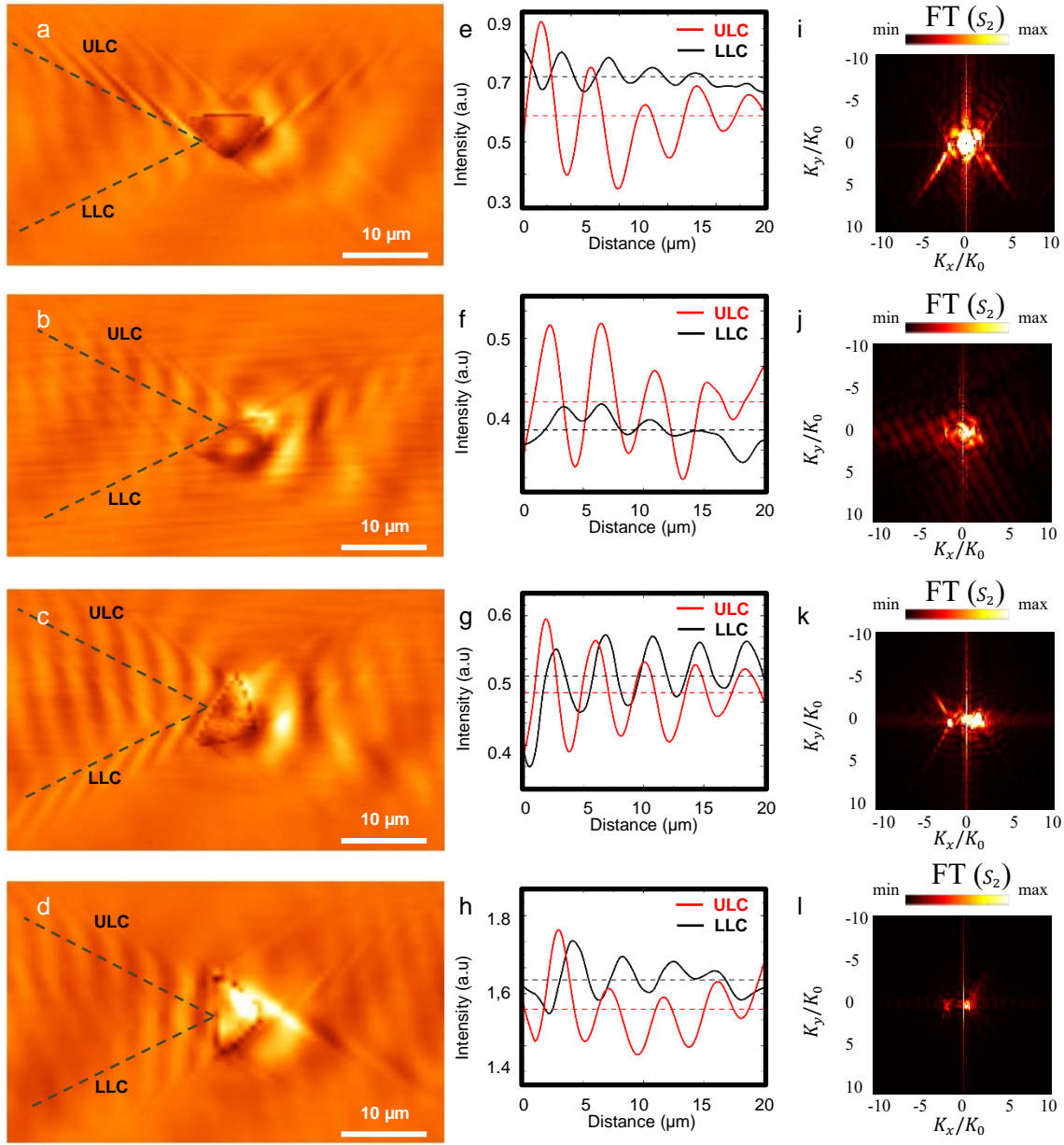

**Figure S4. Determination of directionality of g-HPs for different shapes.** (**a-d**) 50μm × 30μm s-SNOM scan of a triangle-shaped nano antenna marked with the upper line cut (ULC) and lower line cut (LLC) profiles for the rotational angles β = 0°, 30°, 60°, 90°. (**e-h**) Upper and lower line cut intensity profiles of s-SNOM images presented in a-d. (**i-l**) Absolute value of the Fourier transform of the images shown in a-d.

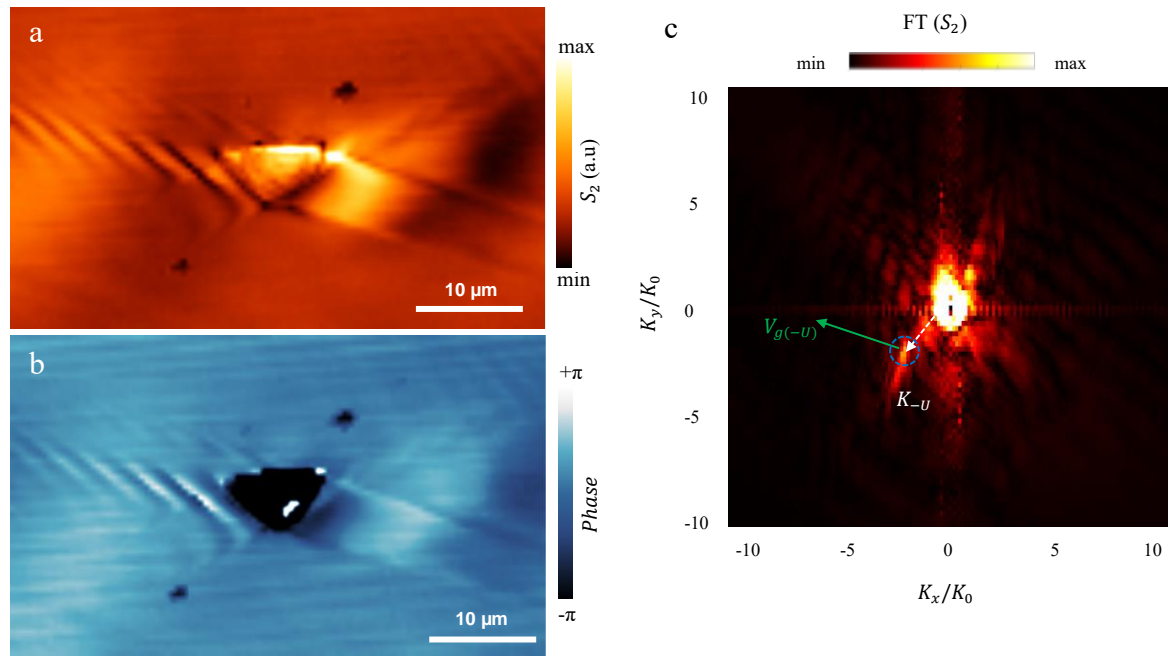

**Figure S5. Fourier Analysis calculation technique of the s-SNOM images.** (**a**) Amplitude signal of experimental near-field images of a triangular shaped antenna-launched g-HPs at the illumination frequency $\omega = 1440\ cm^{-1}$. (**b**) Phase signal of experimental near-field images of a triangular shaped antenna-launched g-HPs at the illumination frequency $\omega = 1440\ cm^{-1}$. (**c**) Fourier transformed $S_2$, generated by the triangular antenna shown in a and b. dotted blue circle highlights the excited mode (bright spot), white arrow indicates the wave vector of the excited mode, and the green arrow indicates the group velocity of the polariton.

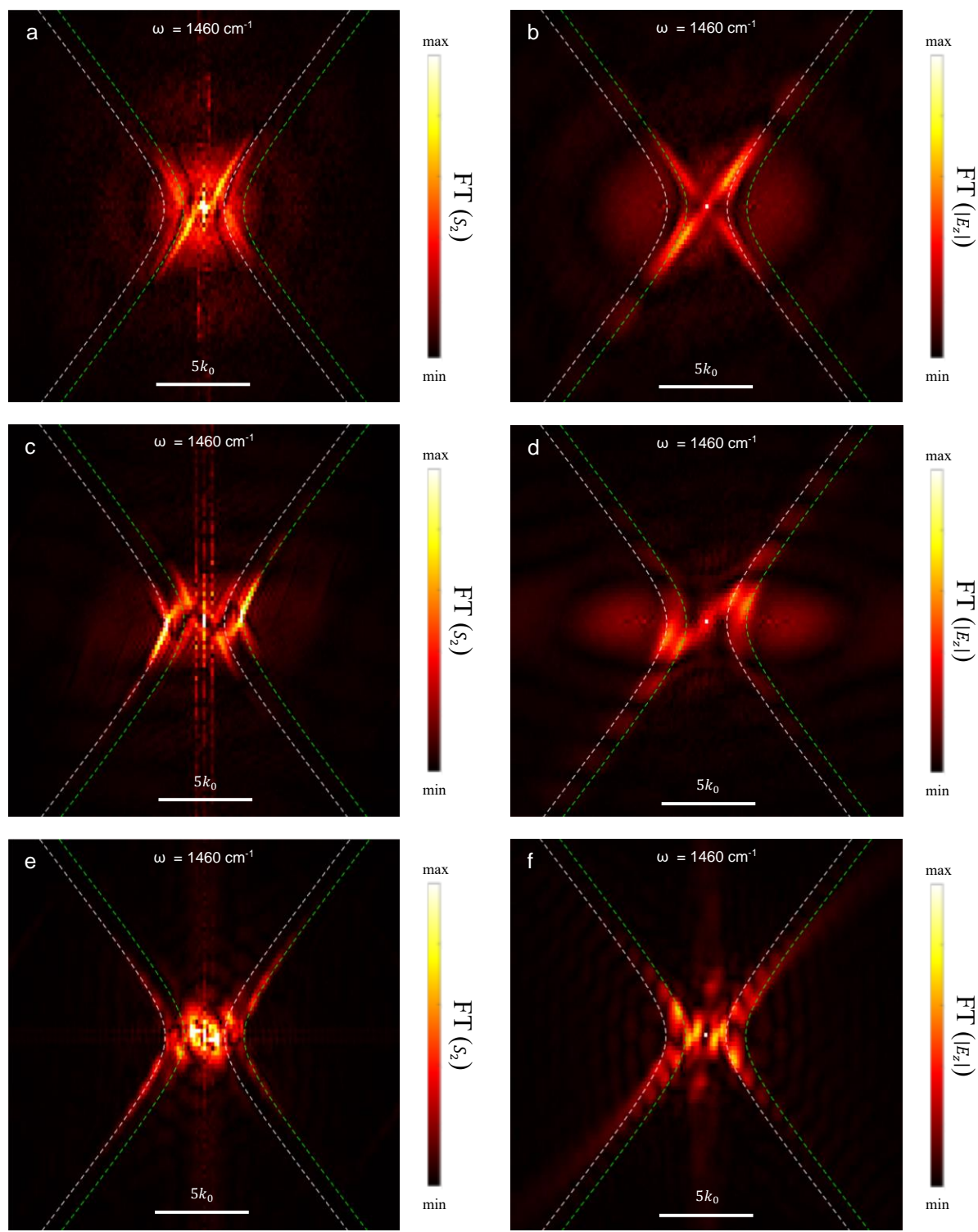

**Figure S6. Fourier transformation analysis of polaritons generated by different shaped nano antennas.** (**a, b**) Fourier transformed $S_2$ and $|E_z|$, generated by the disk antenna shown in figure 1(A and D) in the main text. (**c, d**) Fourier transformed $S_2$ and $|E_z|$, generated by the rectangle antenna shown in figure 1(b and e). (**e, f**) Fourier transformed $S_2$ and $|E_z|$, generated by the triangle antenna showed in figure 1(c, f). The left column corresponds to experimental results, while the right column corresponds to simulation results.

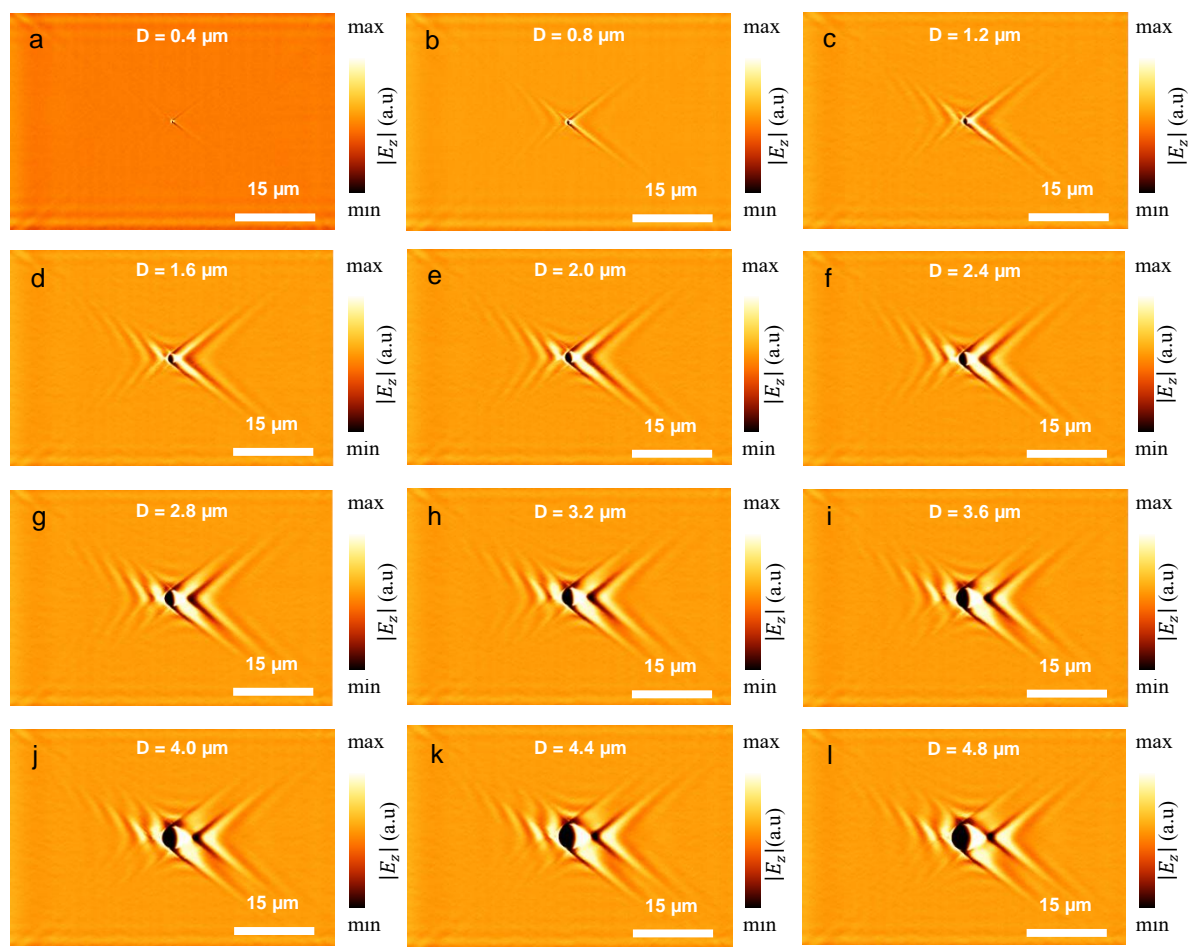

**Figure S7. Polariton propagation length comparison of different sized disk nano antennas.**
**a-l,** simulated near-field images of disk launched hyperbolic polaritons for different sized disk antennas for the illumination frequency 1460 cm$^{-1}$.

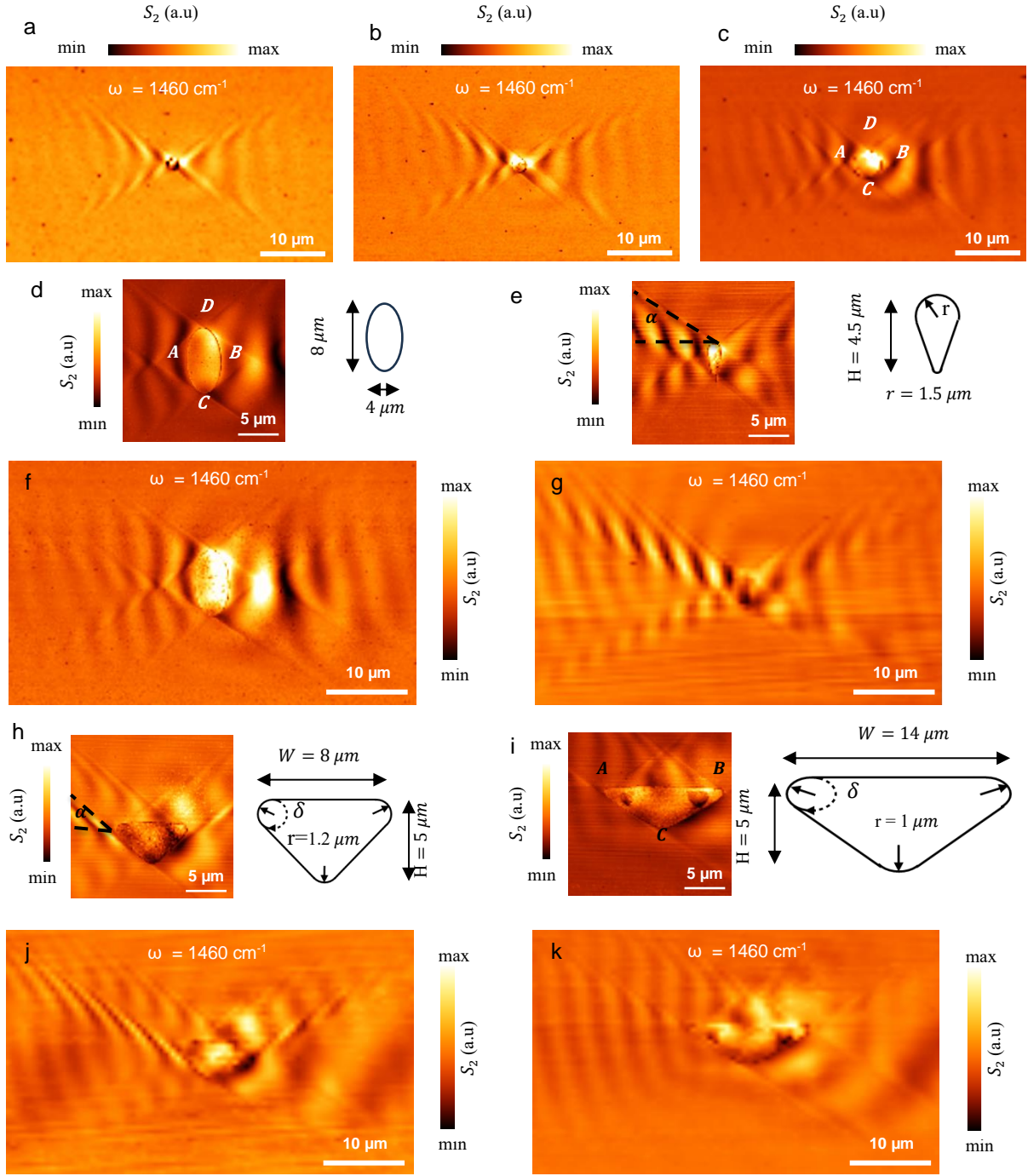

**Figure S8. Polariton excitation behavior from a disk antenna to a triangular antenna based on curvature and edges.** (**a-c**) 50 × 30μm s-SNOM scan of the disk-shaped antennas with a radius of 0.75μm, 1 μm and 2 μm. A-D are the 4 curved points on the edge of the disk-shaped antenna. (**d**) 20 × 20μm near field image and the schematic diagram of an ellipse shaped antenna with a minor axis diameter of 4 μm and a major axis diameter of 8 μm. A-D are the 4 curved

points on the edge of the ellipse-shaped antenna. (**e**) 20 × 20µm near field image and the schematic diagram of a small triangular shaped antenna. Curved point D with a radius of curvature 1.5 µm and a height of 4.5 µm. Open angle (α) indicated in between the two dotted lines. (**f**) 50 × 30µm s-SNOM scan of the ellipse antenna shown in d. (**g**) 50 × 30µm s-SNOM scan of the triangle antenna shown in e. (**h**) 20 × 20µm near field image and the schematic diagram of a triangular shaped antenna. Curved points at A, B and C with a radius of curvatures 1.2 µm and a height of 5 µm. (**i**) 20 × 20µm near field image and the schematic diagram of a small triangular shaped antenna. Curved point at A, B and C with a radius of curvatures 1 µm and a height of 5 µm. Open angle (α) indicated by the two dotted lines. (**j**) 50 × 30µm s-SNOM scan of the triangle antenna shown in h. (**k**) 50 × 30µm s-SNOM scan of the triangle antenna shown in i. δ is the triangular internal angle which represents the inclination of the side of the triangle shape.

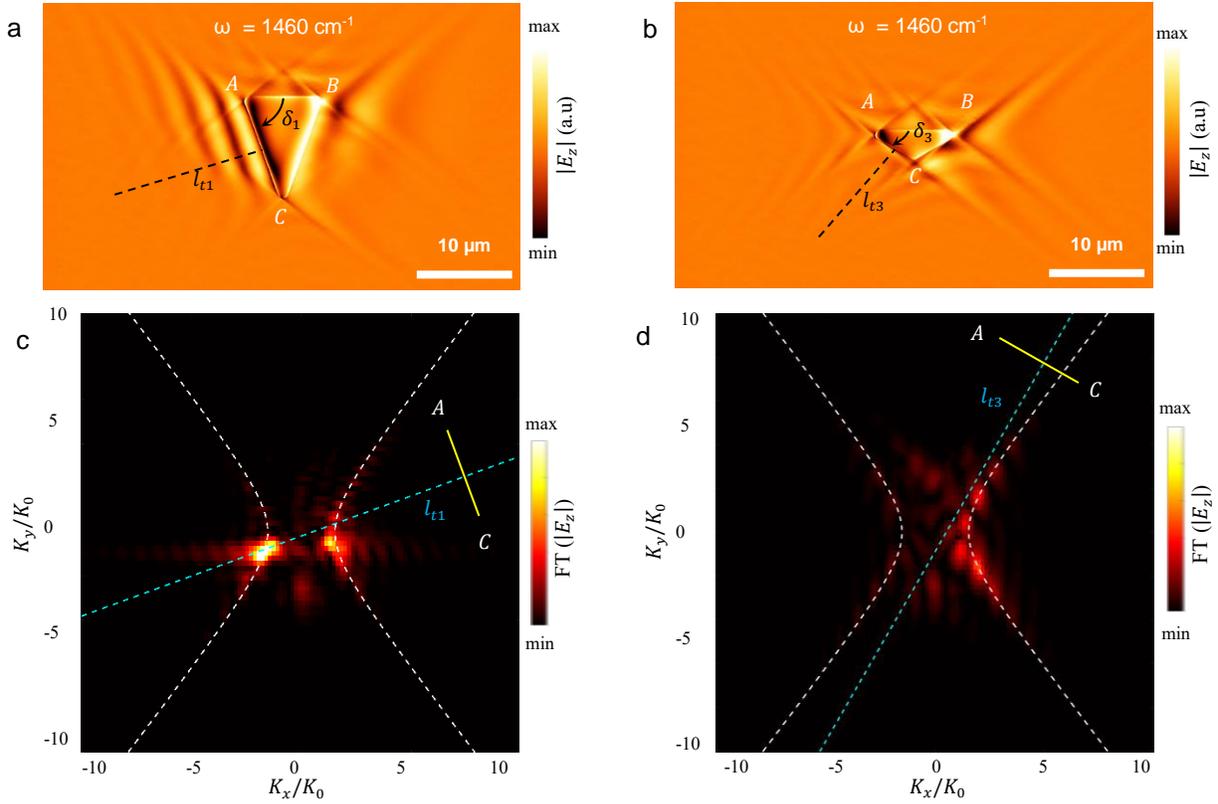

**Figure S9. Effect of negative reflection on directionality for rotated nano antennas.** (**a, b**) Simulated near-field images of antenna-launched g-HPs for triangles with two distinct internal angles (δ). $\delta_1 = 70°$ and $\delta_2 = 30°$. (**c, d**) Absolute value of the Fourier transform of the images a and b. $l_{t1}$ and $l_{t2}$ are the perpendicular axis to the AC physical edge of the triangles with internal angles $\delta_1$ and $\delta_2$.

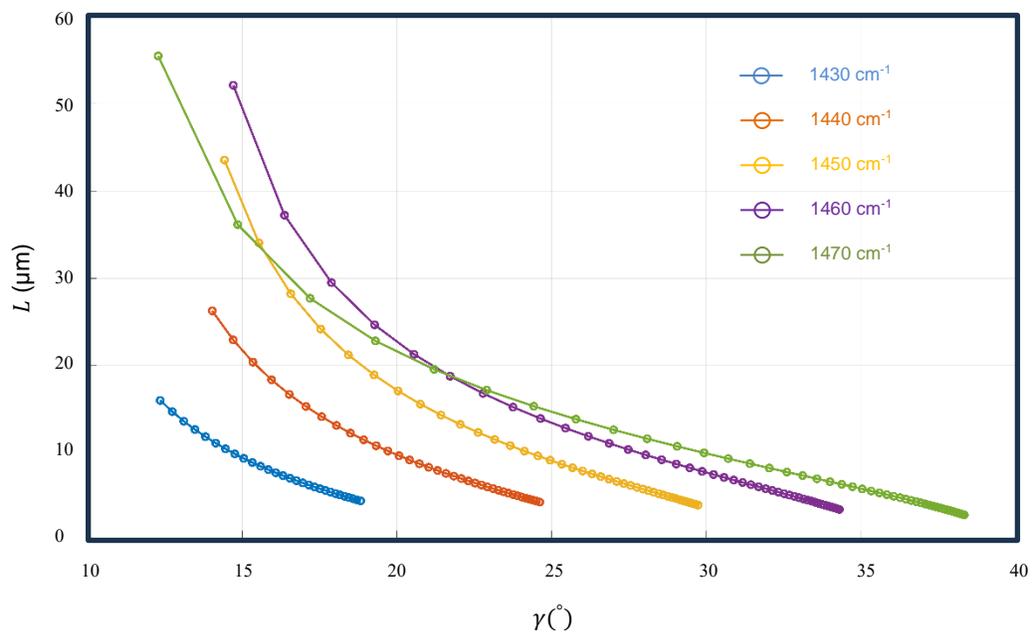

**Figure S10. Analytically calculated propagation length as a function of propagation direction under different excitation frequencies**. Circle and the line represent the calculated data and the line curve connecting the data for the respective excitation frequencies from 1430 cm⁻¹ to 1470 cm⁻¹.

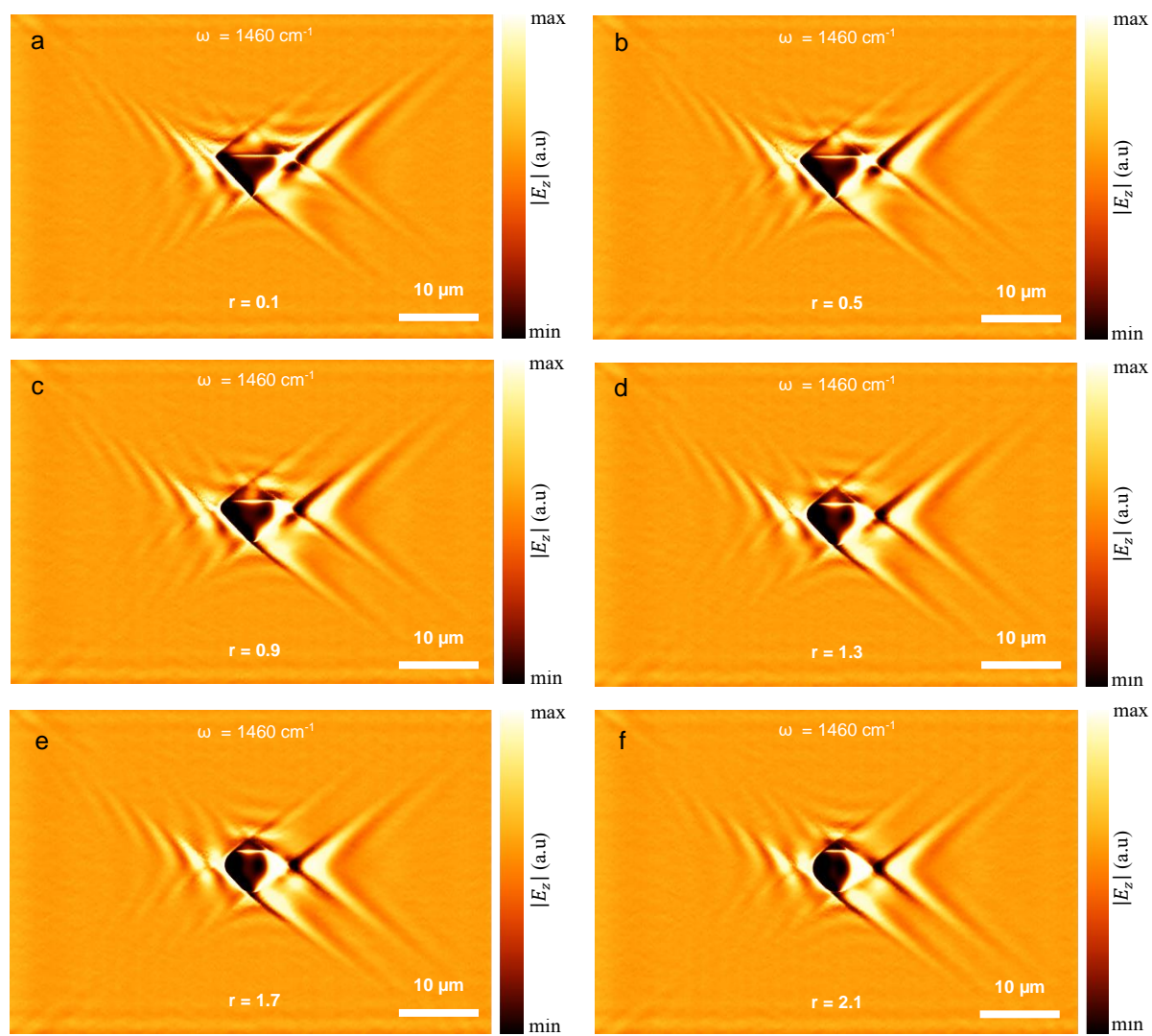

**Figure S11. Polariton propagation length comparison of different shapes. (a-f)** simulated near-field images of triangular shaped nano antennas for different curvatures between r = 0.1 μm to r = 2.5 μm for the illumination frequency of 1460 cm⁻¹.

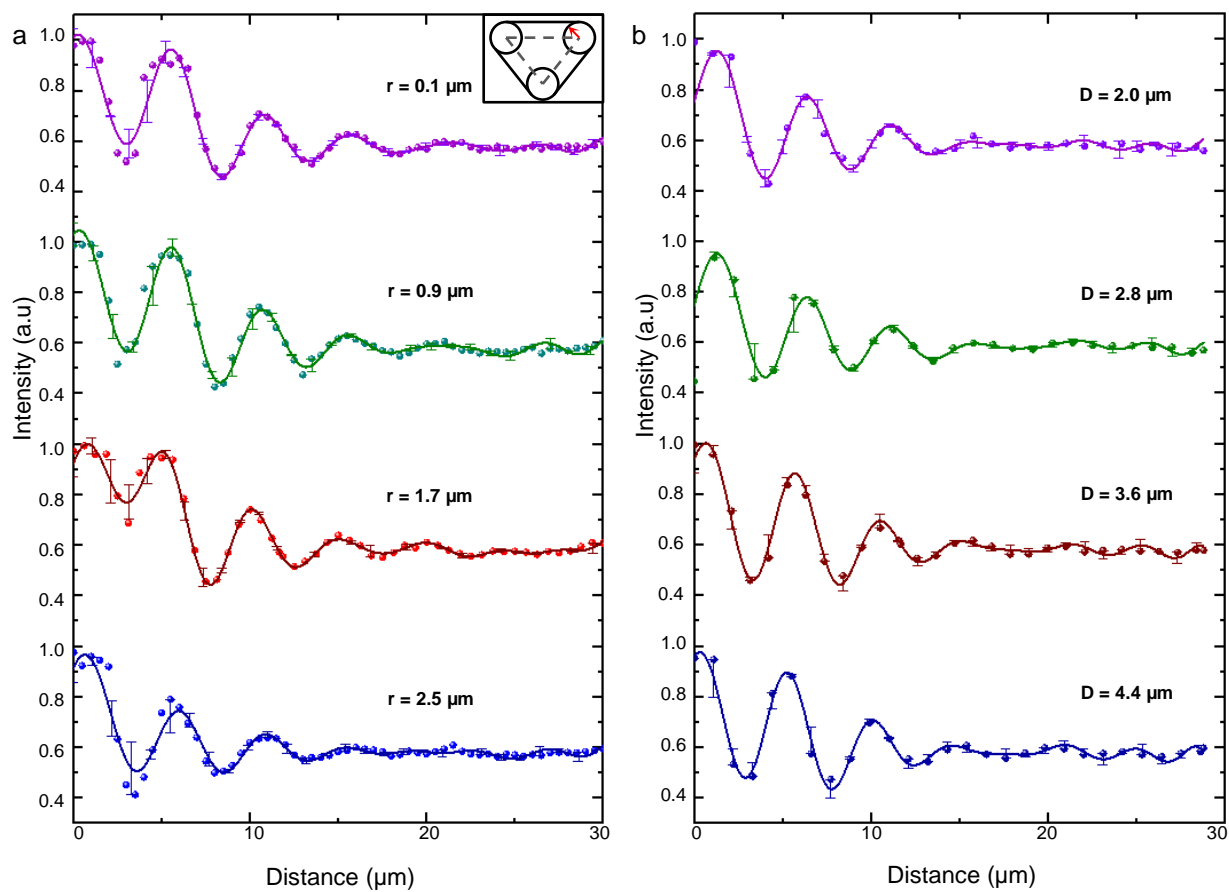

**Figure S12. Variation in the propagation length of g-HPs with design parameters. (a)** Line cut profiles for the different radius of curvatures (r) which converts a triangle to a disk-shaped nano antenna extracted from simulated near-field images of antenna-launched g-HPs. (**b**) Line cut profiles for different diameters of a disk-shape nano antenna extracted from simulated near-field images of antenna-launched g-HPs.